\documentclass[sigconf, table]{acmart}

\AtBeginDocument{%
  }
\AtBeginMaketitle{%
\setcopyright{none}
\copyrightyear{\copyright\ 2025 Copyright held by the owner/author(s).
  Published on \url{OpenProceedings.org} under ISBN 978-3-98318-102-5, series ISSN 2367-2005.
  Distribution of this paper is permitted under the terms of the
  Creative Commons license CC-by-nc-nd 4.0}
\acmYear{2025}
\acmDOI{}
\acmISBN{}

\acmConference[EDBT '26]{Extending Database Technology}{24-27 March 2026}{Tampere (Finland)}
\settopmatter{printacmref=false, printccs=false, printfolios=false}

\newsavebox{\ximagebox}
\newlength{\ximageheight}
\newsavebox{\xglyphbox}
\newlength{\xglyphheight}
\newcommand{\xbox}[1]%
  {\savebox{\ximagebox}{#1}%
  \settoheight{\ximageheight}{\usebox{\ximagebox}}%
  \savebox{\xglyphbox}{\color{white}\char32}%
  \settoheight{\xglyphheight}{\usebox{\xglyphbox}}%
  \raisebox{\ximageheight}[0pt][0pt]{\raisebox{-\xglyphheight}[0pt][0pt]{%
    \makebox[0pt][l]{\usebox{\xglyphbox}}}}%
    \usebox{\ximagebox}%
    \raisebox{0pt}[0pt][0pt]{\makebox[0pt][r]{\usebox{\xglyphbox}}}}

\newsavebox{\LogoBox}
 \sbox{\LogoBox}{\includegraphics[height=1cm]{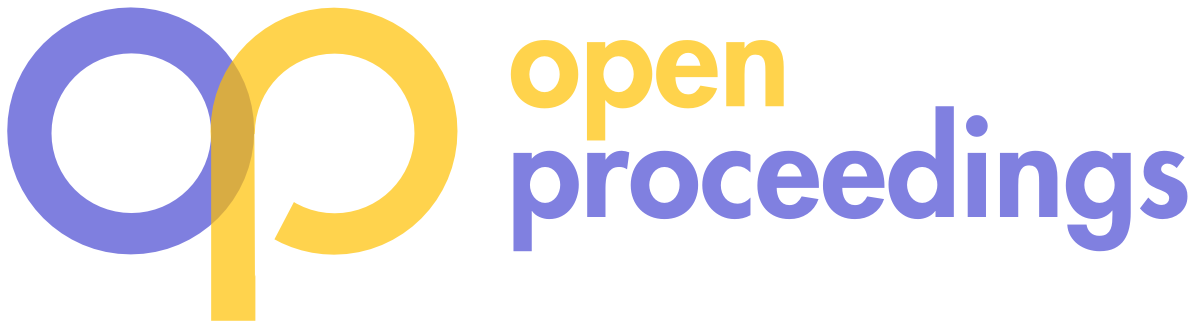}}

\fancypagestyle{firstpagestyle}{%
  \fancyhf{}%
  \fancyfoot{}%
  \fancyhead[L,C]{}
  \fancyhead[R]{\href{https://OpenProceedings.org/}{\raisebox{15pt}[0pt][0pt]{\xbox{\usebox{\LogoBox}}}}}
  }

\hypersetup{%
  pdfcopyright={CC-by-nc-nd}
}

}% \AtBeginMaketitle

\usepackage{amsmath}
	\usepackage{cancel}
\usepackage{mathtools}
\usepackage{graphicx} % can better scale and rotate graphics than graphic package
\usepackage{url}
\usepackage{enumitem} % remove spaces inside 'itemize' [nolistsep]

\usepackage{stmaryrd} % for symbol \llceil
\usepackage{upgreek} % upright Greek letters; together with package "bm" -> upright bold letters
\usepackage{bbding} % for the \CheckmarkBold symbol

\usepackage{listings} % to get sql etc nicely formatted (source code listings)
\usepackage{algcompatible}

\usepackage{tabularx} % table environment that allows to span a whole line
\usepackage{booktabs} % \toprule
\usepackage{multirow}
\usepackage{hhline} % cannot be overwritten by cellcolor
\usepackage{diagbox}
\usepackage{pgfplotstable} % powerful numerical table macros

\usepackage{xcolor,colortbl}    % !!! for some reason, overleaf needs xcolor separately, although that should already have been called by colortbl (10/2019)

\usepackage{tikz} % drawing figures
\usetikzlibrary{matrix}
\usetikzlibrary{calc}
\usetikzlibrary{math}
\usetikzlibrary{arrows.meta,positioning}
\usetikzlibrary{intersections, pgfplots.fillbetween}

\usepackage{easybmat} % \begin{BMAT} alternative syntax for math tabulars and also includes the command addpath which allows to draw a path between matrix elements.

\usepackage{adjustbox}
\def\eox{\unskip\kern 10pt{\unitlength1pt\linethickness{.4pt}$\diamondsuit${}}} % "\eox" command for end of example

\usepackage{rotating}		% for rotated text in tables, \begin{turn}{75}Accuracy\end{turn}

\usepackage{xspace} % for macros

\usepackage{subcaption} % recommended new package for subfigures (package subfig deprecated)
\newcommand{\ak}[1]{{\color{black}{#1}}}
\newcommand{\hide}[1]{}

\usepackage[ruled,noend,linesnumbered]{algorithm2e} % boxruled, linesnumbered,ruled

\usepackage{setspace} % for setting line spacing

\DontPrintSemicolon     % does not print ";" at end of each line
\SetNlSty{}{}{}                % style of line numbers "\Setnlsty{small}{}{:}"
\SetAlgoInsideSkip{smallskip}   % distance between caption rule and beginning of algorithm
\SetAlFnt{\small}			% size of actual algorithm text\scriptsize\sf\rm
\SetAlCapFnt{\small}		% size of name of algorithm
\SetAlCapNameFnt{\small}

\usepackage[capitalise,nameinlink]{cleveref} %nameinlink makes whole word colored

\crefname{section}{Sec.}{Secs.}
\crefname{example}{Ex.}{Exes.}

\hypersetup{                                                            % for both, DVI and PDF
	pdfpagemode=UseNone,               % pdfpagemode=UseOutlines
    pdfstartview=Fit,
    pdfpagelayout=SinglePage,       % pdfpagelayout=SinglePage
    bookmarks=false,
    bookmarksnumbered=true,
    bookmarksopen=false,             % bookmarkstree expanded
    pdftex=true,
    unicode,
    colorlinks=true,       % false: boxed links; true: colored links
    linkcolor=magenta,          % color of internal links (change box color with linkbordercolor)
    citecolor=cyan,  % color of links to bibliography
    filecolor=magenta,      % color of file links
    urlcolor=blue           % color of external links
}

\usepackage{aliascnt}  	
\newaliascnt{corollary}{theorem}

\aliascntresetthe{corollary}

\newaliascnt{example}{theorem}
\newtheorem{example}[example]{Example}
\aliascntresetthe{example}

\newaliascnt{definition}{theorem}
\newtheorem{definition}[definition]{Definition}
\aliascntresetthe{definition}

\newaliascnt{proposition}{theorem}

\aliascntresetthe{proposition}

\newaliascnt{lemma}{theorem}

\aliascntresetthe{lemma}

\newaliascnt{conjecture}{theorem}

\aliascntresetthe{conjecture}

\newtheorem{questionW}{Question}
\newtheorem{resultW}{Result}

{\end{itshape}
}
\AtEndPreamble{%
    \hypersetup{colorlinks,
      linkcolor=purple,
      citecolor=blue,
      urlcolor=ACMDarkBlue,
      filecolor=ACMDarkBlue}}

\usepackage{tcolorbox}		% also allows to have shaded backgrounds around environments
\tcbuselibrary{breakable,skins}		% allow to break boxes, allow to use "enhanced jigsaw"

\tcbset{examplestyle/.style={
		enhanced jigsaw,	% if box is broken, don't show border at continuation
		colback=blue!08,	% !20	gray!10
		colframe=blue!08,	% !40
		arc=2mm,
		boxrule=0pt,		% 0pt
		left=1mm,
		right=1mm,
		left skip= 0mm,  % breaks ACM template unless specified for specific style (4/2019)
		right skip= 0mm, % breaks ACM template unless specified for specific style (4/2019)
		top=1mm,		% 0mm, -1mm
		bottom=1mm,		% problem in package with eq at end of env. just add: \tcbset{bottom=2mm} \tcbset{bottom=0mm}
		breakable,		% allows page break
		parbox = false,		% restores normal text behavior. ELSE: paragraphs are formatted slightly different as the main text. DOWNSIDE: unwanted side effects
		before={\par\pagebreak[0]\vspace{1mm}\parindent=0pt},		
		after={\par\pagebreak[0]\vspace{1mm}\parindent=0pt},				
		bottomrule = 0mm,
		boxsep = 0mm,					% common padding of hlengthi between the text content and the frame of the box, added to left, right, top, ...
		topsep at break=0pt,			% Additional vertical space at the top of middle and last parts in a break sequence
		bottomsep at break=0pt,			% Additional vertical space at the first and middle parts in a break sequence		
		pad at break=0mm,
		pad before break=1mm,		
		pad after break=1mm,		
		bottomrule at break=0mm,
		toprule at break=0mm,		
		}}

\usepackage{footnote}

\tcolorboxenvironment{example}{examplestyle}
\newcommand{\resultbox}[1]{
\begin{tcolorbox}[
	enhanced jigsaw,		% if box is broken, don't show border at continuation
	colback=red!5,
	colframe=red!75!black,	
	arc=0mm,
	left skip=-1mm,
	right skip=-1mm,	
	left=0mm,
	topsep at break=1mm,			% Additional vertical space at the top of middle and last parts in a break sequence
	right=0mm,
	top=0mm,
	bottom=0mm,		% problem in package with eq at end of env. justadd:\tcbset{bottom=2mm}\tcbset{bottom=0mm}
	breakable,		% allows page break
	parbox = false		% restores normal text behavior. ELSE: paragraphs are formatted slightly different as the main text. DOWNSIDE: unwanted side effects
]
\emph{#1}
\end{tcolorbox}
}

\makeatletter
\DeclareRobustCommand*\uell{\mathpalette\@uell\relax}
\newcommand*\@uell[2]{
  \setbox0=\hbox{$#1\ell$}
  \setbox1=\hbox{\rotatebox{10}{$#1\ell$}}
  \dimen0=\wd0 \advance\dimen0 by -\wd1 \divide\dimen0 by 2
  \mathord{\lower 0.1ex \hbox{\kern\dimen0\unhbox1\kern\dimen0}}
}
\makeatother

\usepackage{marginnote}

\newcommand{\introparagraph}[1]{\textbf{#1.}} % define own new subsection type: noindent, bold (textsc)

\newcommand{\ourmethod}{\textsc{DUST}\xspace}
\renewcommand{\epsilon}{\varepsilon} % nicer epsilon symbol
\usepackage{scalerel}
\usepackage{tikz}
\usetikzlibrary{svg.path}

\definecolor{orcidlogocol}{HTML}{A6CE39}
\tikzset{
  orcidlogo/.pic={
    \fill[orcidlogocol] svg{M256,128c0,70.7-57.3,128-128,128C57.3,256,0,198.7,0,128C0,57.3,57.3,0,128,0C198.7,0,256,57.3,256,128z};
    \fill[white] svg{M86.3,186.2H70.9V79.1h15.4v48.4V186.2z}
                 svg{M108.9,79.1h41.6c39.6,0,57,28.3,57,53.6c0,27.5-21.5,53.6-56.8,53.6h-41.8V79.1z M124.3,172.4h24.5c34.9,0,42.9-26.5,42.9-39.7c0-21.5-13.7-39.7-43.7-39.7h-23.7V172.4z}
                 svg{M88.7,56.8c0,5.5-4.5,10.1-10.1,10.1c-5.6,0-10.1-4.6-10.1-10.1c0-5.6,4.5-10.1,10.1-10.1C84.2,46.7,88.7,51.3,88.7,56.8z};
  }
}

\RequirePackage{etoolbox}
\DeclareRobustCommand\orcidicon[1]{\href{https://orcid.org/#1}{\mbox{\scalerel*{
\begin{tikzpicture}[yscale=-1,transform shape]
\pic{orcidlogo};
\end{tikzpicture}
}{|}}}} % includes various useful macros (keep macros separate from main content files)
\usepackage{array}
\usepackage{makecell}

\newcommand{\cc}{olive}
\newcommand{\algocomment}[1]{\textcolor{\cc}{{//#1}}}

\newcommand{\topk}{top-\textit{k}\xspace}

\newif\ifreport
\reporttrue  % Uncomment for Technical Report
\newcommand{\paperorreport}[2]{\ifreport #2\else #1\fi}

\newif\ifhidedata
\hidedatatrue

\begin{document}

%%
%% The "title" command has an optional parameter,
%% allowing the author to define a "short title" to be used in page headers.
\paperorreport{\title{Diverse Unionable Tuple Search: \\Novelty-Driven Discovery in Data Lakes}}{\title{Diverse Unionable Tuple Search: %A 
Novelty-Driven Discovery in Data Lakes [Technical Report]}}

\author{Aamod Khatiwada}
 \orcid{0000-0001-5720-1207}
\affiliation{%
  \institution{Northeastern University}
  %\streetaddress{P.O. Box 1212}
  \city{Boston}
  \state{Massachusetts}
  \country{USA}
  %\postcode{43017-6221}
}
%\email{trovato@corporation.com}
% \email{khatiwada.a@northeastern.edu}
\email{khatiwadaaamod@gmail.com}

\author{Roee Shraga}
\orcid{0000-0001-8803-8481}
\affiliation{%
  \institution{Worcester Polytechnic Institute}
  \city{Worcester}
  \state{Massachusetts}
  \country{USA}
  %\streetaddress{P.O. Box 1212}
  %\city{Dublin}
  %\state{Ireland}
  %\postcode{43017-6221}
}
\email{rshraga@wpi.edu}

\author{Ren\'ee J. Miller}
\orcid{0000-0002-1484-4787}
\affiliation{%
  \institution{U. Waterloo \& Northeastern U.}
  \city{Waterloo}
  \state{Ontario}
  \country{Canada}
  }
%\email{trovato@corporation.com}
\email{rjmiller@uwaterloo.ca}
%%
%% By default, the full list of authors will be used in the page
%% headers. Often, this list is too long, and will overlap
%% other information printed in the page headers. This command allows
%% the author to define a more concise list
%% of authors' names for this purpose.
% \renewcommand{\shortauthors}{Aamod Khatiwada et al.}
\renewcommand{\shortauthors}{Aamod Khatiwada, Roee Shraga, \& Renée J. Miller}

%%
%% The abstract is a short summary of the work to be presented in the
%% article.
% \input{sections/cover_letter}
\begin{abstract}
Unionable table search techniques input a query table from a user and search for data lake tables that can contribute additional rows to the query table. The definition of unionability is generally based on similarity measures which may include similarity between columns (e.g., value overlap or semantic similarity of the values in the columns) or tables (e.g., similarity of table embeddings). Due to this and the large redundancy in many data lakes (which can contain many copies and versions of the same table), the {\bf most} unionable tables may be identical or nearly identical to the query table and may contain little new information. Hence, we introduce the problem of identifying unionable tuples from a data lake that are diverse with respect to the tuples already present in a query table. We perform an extensive experimental analysis of well-known diversity algorithms applied to this novel problem and identify a gap that we address with a novel, clustering-based tuple diversity algorithm called \ourmethod.
\ourmethod uses a novel embedding model to represent unionable tuples that outperforms other tuple representation models by at least $15\%$ when representing unionable tuples. Using real data lake benchmarks, we show that our diversification algorithm is more than six times faster than the most efficient diversification baseline.  We also show that it is more effective in diversifying unionable tuples than existing diversification algorithms.
\end{abstract}

%%
%% The code below is generated by the tool at http://dl.acm.org/ccs.cfm.
%% Please copy and paste the code instead of the example below.
%%
%% A "teaser" image appears between the author and affiliation
%% information and the body of the document, and typically spans the
%% page.
% \begin{teaserfigure}
%   \includegraphics[width=\textwidth]{sampleteaser}
%   \caption{Seattle Mariners at Spring Training, 2010.}
%   \Description{Enjoying the baseball game from the third-base
%   seats. Ichiro Suzuki preparing to bat.}
%   \label{fig:teaser}
% \end{teaserfigure}

%\received{13 January 2023}
%\received[revised]{12 March 2009}
%\received[accepted]{5 June 2009}

%%
%% This command processes the author and affiliation and title
%% information and builds the first part of the formatted document.
\maketitle
\setcounter{page}{1}	% reset page numbering to 1

% \pagestyle{\vldbpagestyle}
% \begingroup\small\noindent\raggedright\textbf{PVLDB Reference Format:}\\
% \vldbauthors. \vldbtitle. PVLDB, \vldbvolume(\vldbissue): \vldbpages, \vldbyear.\\
% \href{https://doi.org/\vldbdoi}{doi:\vldbdoi}
% \endgroup
% \begingroup
% \renewcommand\thefootnote{}\footnote{\noindent
% This work is licensed under the Creative Commons BY-NC-ND 4.0 International License. Visit \url{https://creativecommons.org/licenses/by-nc-nd/4.0/} to view a copy of this license. For any use beyond those covered by this license, obtain permission by emailing \href{mailto:info@vldb.org}{info@vldb.org}. Copyright is held by the owner/author(s). Publication rights licensed to the VLDB Endowment. \\
% \raggedright Proceedings of the VLDB Endowment, Vol. \vldbvolume, No. \vldbissue\ %
% ISSN 2150-8097. \\
% \href{https://doi.org/\vldbdoi}{doi:\vldbdoi} \\
% }\addtocounter{footnote}{-1}\endgroup
%%% VLDB block end %%%

%%% do not modify the following VLDB block %%
%%% VLDB block start %%%
% \ifdefempty{\vldbavailabilityurl}{}{
% \vspace{.3cm}
% \begingroup\small\noindent\raggedright\textbf{PVLDB Artifact Availability:}\\
% The source code, data, and/or other artifacts have been made available at \url{\vldbavailabilityurl}.
% \endgroup
% }
\section{Introduction}
\label{section:introduction}

\begin{figure*}[ht]
  \includegraphics[scale = 0.5]{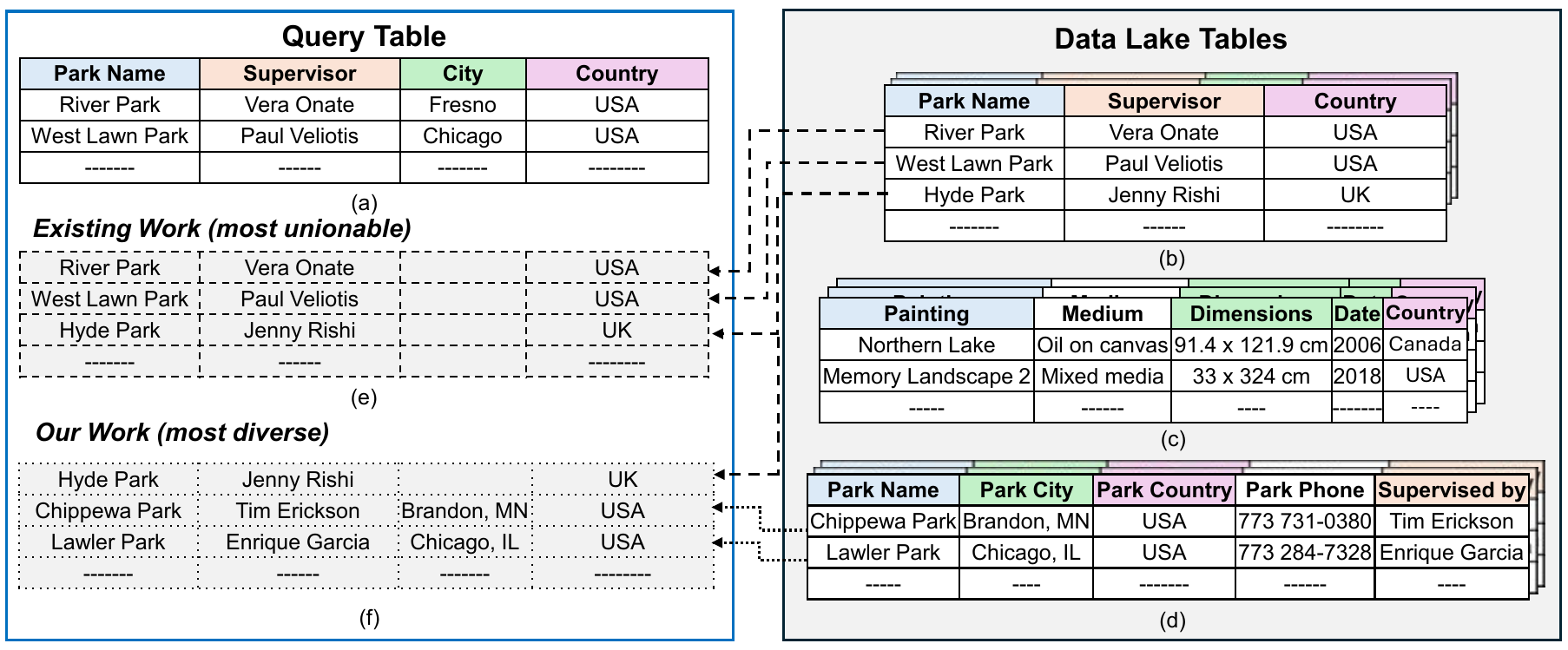}
  \caption{Table (a) is a Query Table (left) and Tables (b)-(d) are Data Lake Tables (right). Existing Work will add the tuples in Table (e) (most unionable) while our work will return the ones in Table (f), which are also the most diverse. 
  }
\label{fig:park_tables}
\end{figure*}

Content-based table search (or table as a query) has become the dominant paradigm for table discovery in data lakes~\cite{2023_fan_table_discovery_tutorial,2023_dong_deepjoin, 2023_hu_autotus, 2023_hai_datalake_survey}, particularly when metadata may be missing, ambiguous, inconsistent, or incomplete~\cite{2019_nargesian_data_lake_management, 2023_khatiwada_dialite}.  
In this paradigm, users input a table as a query, and the table search system uses the data available within the query table to retrieve 
relevant tables. 
Among different types of relevance search~\cite{2023_hai_datalake_survey, 2023_fan_table_discovery_tutorial, khatiwada2024tabsketchfm}, we focus on \emph{Table Union Search}~\cite{2018_nargesian_tus}, where the objective is to search for additional \emph{"unionable"} tables. %that add new rows (tuples) to the query table. 
Note that an important end goal of searching for the unionable tables is to add new rows to a given query table.
However, table union search techniques return the
data lake tables that are most unionable to the query table~\cite{2023_khatiwada_santos, 2018_nargesian_tus, 2020_bogatu_d3l}. They measure unionability between tables based on different similarity measures including value overlap between the columns~\cite{2018_nargesian_tus, 2020_bogatu_d3l}, knowledge graphs concept similarity~\cite{2018_nargesian_tus, 2023_khatiwada_santos}, word embedding similarity~\cite{2018_nargesian_tus}, 
similarity of table representations~\cite{2023_fan_starmie, 2023_hu_autotus}, and so on. 
But such similarity-based techniques tend to return tables containing similar or even redundant tuples (with respect to the query table) in the top rankings. In fact, a table is most similar to itself.

\begin{example}
\label{example:union_search}
% For instance, c
Consider Query Table (a) and Data Lake Tables (b), (c) and (d) in~\cref{fig:park_tables}. Tables (a), (b) and (d) are about parks and their different properties, and both Tables (b) and (d) are unionable with the query table. Table (c) is about paintings and as it shares only the country attribute with (a), it is determined to be not unionable with the query table. 
Since table union search techniques~\cite{2023_hu_autotus, 2023_khatiwada_santos, 2018_nargesian_tus,2023_fan_starmie} 
measure similarity between query and data lake tables to infer unionability, they return Table (b) as the most unionable table %with the query table 
(Table (d) while unionable is less similar to the query). However, 
Table (b) 
is mostly a copy of the query table with a single new tuple
and thus does not extend the user's analysis much (we illustrate this in Table (e) which is "most unionable").
In contrast, Table (d) is both unionable and contains tuples with new information.
\end{example}

% \revision{
If the data lake contains tables with overlapping tuples, these are likely to be returned by a union search method.
Others have documented the tremendous redundancy in real data lakes~\cite{2019_zhu_josie}, noting
that almost 90\% of the data found in newly created datasets duplicates information already present in existing datasets in a lake~\cite{2023_shraga_explain_dv}. 
So, the top-ranked tables returned by a similarity-based union search technique may 
not offer much new information.
An example of the impact of such redundancy could be if one were trying to integrate the discovered tables 
to achieve a fairness goal~\cite{2022_nargesian_responsible_data_integration, 2019_yang_diveristy_constraints, 2018_stoyanovich_online_set_selection} where most approaches assume that the discovered tables  have enough tuples available to satisfy fairness constraints~\cite{2022_nargesian_responsible_data_integration}. However, if redundancy is high, this may not be the case.
Table search is used to find training data for machine learning models~\cite{2023_galhotra_metam, 2023_han_datacentric_ml} and redundancy in the discovered data may hamper the model's ability to generalize.  
A user study on diversifying information retrieval (IR) search results reflects that when looking for new information,  users are interested in finding more information within their specific subtopics of interest, rather than arbitrarily looking for any new information~\cite{2008_xu_diversity_user_study}. 
In our case, for instance, a user is looking for 
information on new parks that are not present in the query table. 
% }
% \revision{
In this paper, we present a novel evaluation of the ability of existing diversity algorithm to diversify the tuples returned by unionability methods.
We further offer \ourmethod, a method designed to address gaps identified in existing algorithms and scalably provide a diverse set of unionable tuples. 
% }  
Referring to Example~\ref{example:union_search} (\cref{fig:park_tables}), our proposed approach will find unionable tuples from a set of unionable tables where the tuples add new and diverse information to the query table as illustrated by Table (f).

\begin{figure}[h]
  \includegraphics[width=.21\textwidth]{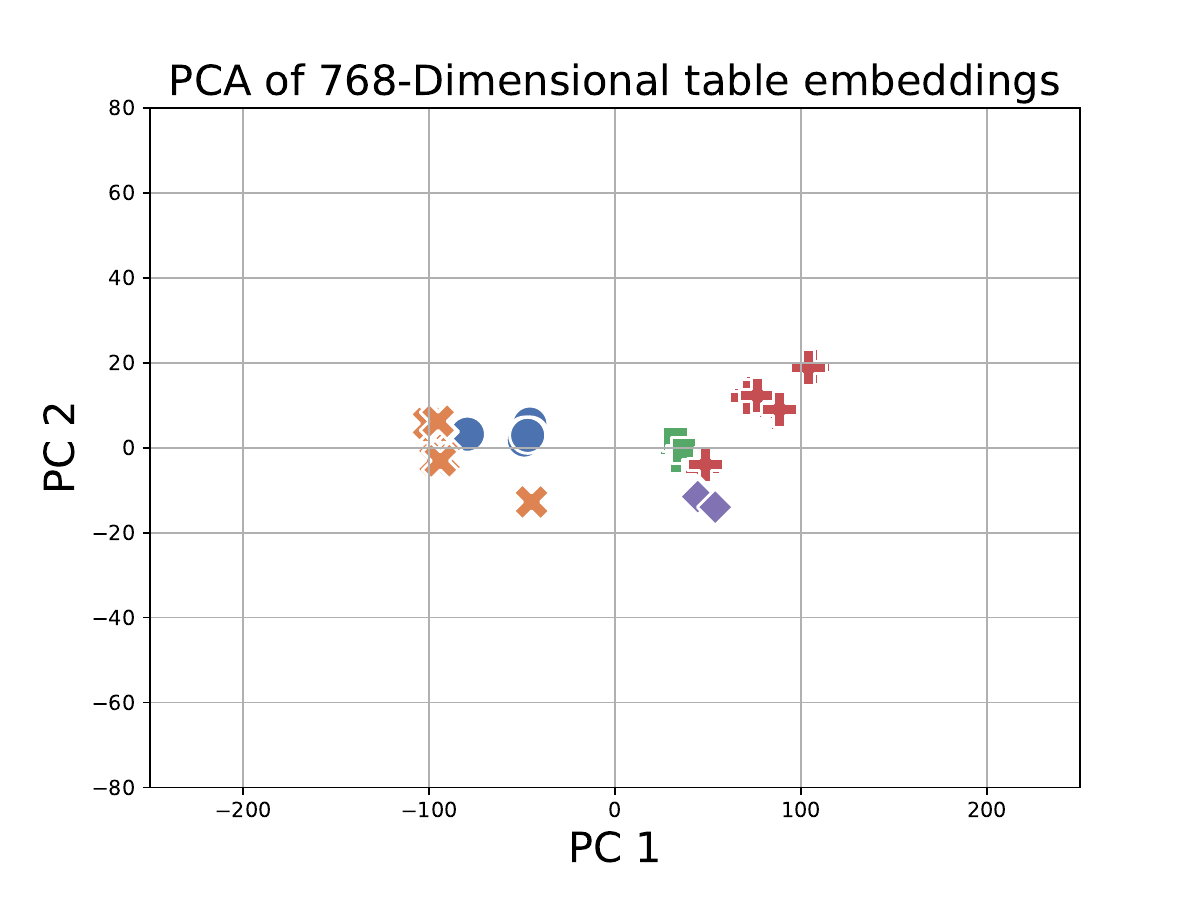}
  \includegraphics[width=.195\textwidth]{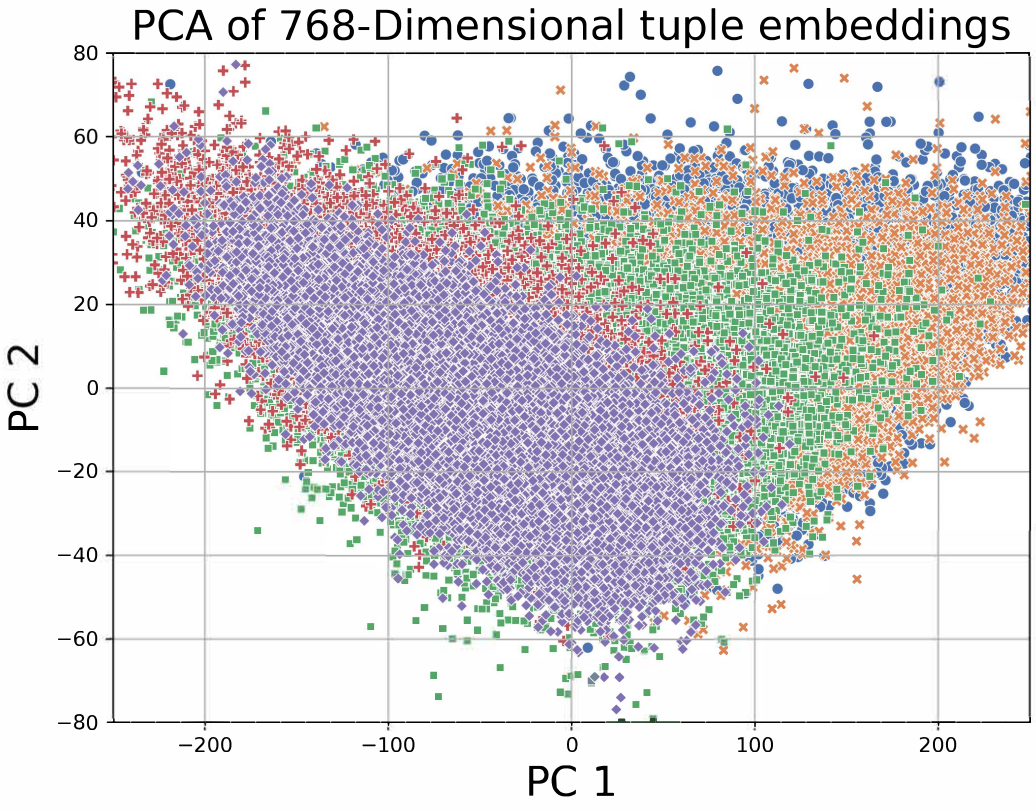}
  \caption{Table (left) and Tuple (right) embedding distributions of unionable tables (tuples) from Open Data~\cite{2023_khatiwada_santos}. 
    Each color/marker represents a distinct set of  unionable tuples/tables. Tables from different sets are non-unionable.
  }
\label{fig:motivation_embeddings}
\end{figure}

% \revision{
\introparagraph{Tuple vs.~Table Diversity}
Our important design decision 
is to return a set of diverse, unionable \textit{tuples} (which may come from different tables), rather than a set of diverse, unionable \textit{tables}. 
\cref{fig:motivation_embeddings} provides a motivation for this decision.   Using any tuple or table embedding technique, we can plot (using PCA) a two-dimensional representation of the embedding similarity among unionable and non-unionable tuples or tables.
\cref{fig:motivation_embeddings} does this using embeddings we will introduce in \cref{section:tuple_representation}.
On the left, we plot embeddings of five sets (each set in a different color/marker) of unionable tables from the SANTOS benchmark which contains tables from Open Data~\cite{2023_khatiwada_santos}.  Tables from different sets are non-unionable.  
While there is some clustering structure in the plot, the unionable tables are not terribly diverse in the embedding space and even some non-unionable tables have quite similar representations.
This shows that the diversity of the tables is limited. %}
The right side contains the same plot for unionable and non-unionable tuples.
Tuples are spread around the space much more, and even unionable tuples (with same color/marker) can appear very far apart in the embedding space. 
This suggests that diversifying unionable tables may have limited effect whereas selecting a diverse set of tuples from a set of unionable tables may well achieve our goal of providing new and diverse search results.  In addition,
directly searching for diverse unionable tuples may be infeasible
since it requires an index over all tuples in a lake (rather than tables).
% }

Extensive IR research%in Information Retrieval (IR)
, particularly in recommendation systems and web search, has also focused on the diversification of search results~\cite{2005_adomavicius_recommendation_survey, 2014_vargas_diversity_recommender, 2020_gao_toward_fairer_search_results}.
% \revision{
While IR solutions, at large, address very large repositories of data (e.g., documents), when it comes to diversification, they typically focus on smaller sets~\cite{2014_vargas_diversity_recommender, 2018_cecchini_topic_retrieval, 2020_gao_toward_fairer_search_results}. Diversification, in this case, is applied as a second step over human-consumable sets of information such as web pages or items on an e-commerce website.
% }
% \revision{
Accordingly, they experiment with relatively small sets
%number of diverse output results~
(e.g., a prior work considers relevance item set size up to 5000 and diverse output result size up to 35~\cite{2011_vieira_divdb}; and a popularly used web search dataset has a set of queries each with 100 relevant google web search results among which a small (typically, 5 or 10 items out of the 100) diverse set can be extracted~\cite{2020_gao_toward_fairer_search_results}).
% }
%However, i
In our scenario, the challenge lies in diversifying large sets  (tens of thousands) of unionable tuples, which are intended to be processed in automated pipelines for downstream tasks~\cite{2023_galhotra_metam}.

Our method \ourmethod (Diverse Unionable Tuple Search)~\cite{2026_khatiwada_dust}, which given a query table, searches for unionable 
tables
from a
data lake, and post-processes them to output k-diverse tuples that 
are diverse with respect to the query table and each other.
We do not assume a small set of retrieved items and we do not assume a small $k$ as in information retrieval, rather
\ourmethod considers outputting (a potentially large) set of diverse tuples from a larger set of unionable tuples.
%We use a new 
We further fine-tune a transformer-based model that represents the unionable tuples in an embedding space. 
% \revision{
We evaluate unionable table search techniques, emphasizing their effectiveness in achieving diversification. We further adopt and combine ideas from diversification and table union search to present a new algorithm. 
% }  
Specifically, our contributions are:\\
% \textbf{Contributions:}
% \begin{itemize}
    % \item 
    \noindent\introparagraph{$\bullet$ Diverse Unionable Tuple Search} 
%    For the first time in the literature, w
We  
identify and address
the new problem of searching for diverse tuples from a data lake that can be unioned with the given query table. %We also propose the \ourmethod system as a solution to the problem.

    \noindent\introparagraph{$\bullet$ Tuple Diversification Algorithm} We present \ourmethod, a new algorithm to diversify unionable data lake tuples with respect to the query table.
    The algorithm also uses a %novel 
    transformer-based fine-tuned model for tuple representation.
    
    \noindent\introparagraph{$\bullet$ Extensive Empirical Evaluation}
    We run experiments to show the effectiveness and efficiency of \ourmethod in searching for diverse unionable tuples. Specifically, our novel \ourmethod model outperforms baseline tuple embedding models by at least 15\% in terms of accuracy, 
    when used to represent unionable tuples.
    
         Furthermore, we show that \ourmethod's tuple diversification algorithm is over six times faster than the most effective baseline~\cite{2011_vieira_divdb}. Nevertheless, in most cases, our algorithm is more effective than all the baselines when diversifying tuples from existing table union search  benchmarks~\cite{2023_khatiwada_santos, 2023_pal_ugen}. We further provide a case-study intuitively illustrating the practical benefits of the presented algorithm.

    %\item 
    
    \noindent\introparagraph{$\bullet$ Open-source Code and Data}
    We make all our resources open-source: 
    \url{https://github.com/northeastern-datalab/dust}

\section{Related Work}
\label{section:related_work}

To the best of our knowledge, no %system considers searching for a 
prior work searches for a diverse set of unionable tuples from a data lake. So, we cover related work on finding unionable tables and on diversifying search results.

\introparagraph{Table Union Search}
\citet{2018_nargesian_tus} introduced the problem of searching for top-k unionable tables from a data lake, given a query table. They consider a data lake table to be unionable with a query table if they share unionable columns. 
To determine unionable columns, they defined three statistical tests based on value overlap, knowledge graph class overlap, and word embedding similarity between the columns.  \citet{DBLP:journals/pvldb/ZhuNPM16} introduced the joinable table search problem with numerous extensions~\cite{2019_zhu_josie,2023_dong_deepjoin} and \citet{2020_bogatu_d3l} presented $\mathbf{D^3L}$  that searches for data lake tables that are related (meaning either unionable or joinable). 
\citet{2023_khatiwada_santos} introduced \textbf{SANTOS} that improves table union search accuracy by considering column semantics as well as the semantics of binary relationships between column pairs. In \textbf{Starmie}, \citet{2023_fan_starmie} proposed a contrastive-learning framework that instead of looking at columns individually or looking at the binary relationships between the column pairs, captures the context of the entire table in the form of contextualized column embeddings and uses such embeddings to improve table union search accuracy. Hu et al.~\cite{2023_hu_autotus} proposed \textbf{AutoTUS} where they contextualized relationships between the column pairs to capture table contexts and use them to find  unionable tables from the data lakes. 
~\citet{2023_mirzaei_tus_preference} explore the search for similar data lake tables based on user preferences, incorporating diversity among their four major preference criteria. While their primary focus is on finding tables with new sets of columns, we concentrate on the addition of new tuples. Nevertheless, they utilize an existing diversification algorithm from prior work~\cite{2011_vieira_divdb} that we use as a baseline.
In summary, existing techniques return top-k unionable tables whereas, we output k unionable and diverse tuples for a given query table.

\introparagraph{Search result diversification}
The information retrieval literature
considers diversifying search results (aka novelty search)~\cite{1998_carbonell_mmr_diversity, 2005_adomavicius_recommendation_survey, 2014_vargas_diversity_recommender}.
They generally input a relevant item set to a query and output a subset of items that maximize a diversity score given by a diversification function~\cite{2017_zheng_query_diversification_survery}. 
% \revision{
The diversification function is defined by considering a trade-off between relevance and diversity that is controlled using a user-provided parameter, where increasing relevance decreases diversity and vice-versa. Furthermore, they are based on different objectives such as maximizing the sum of distances between the selected items (max-sum diversification)~\cite{2011_vieira_divdb, 1998_carbonell_mmr_diversity, 2003_jain_knn_diversity, 2017_borodin_max_sum_diversification}, maximizing the minimum distance between the selected items (max-min diversification)~\cite{2021_moumoulidou_diversity_fairness}, retrieving items in a diverse set having distances greater than a given threshold (threshold-based diversification)~\cite{2015_drosou_disc_diversity}, and more.
% }
% \revision{
Note that selecting k-diverse items from a given set is an expensive problem~\cite{2011_vieira_divdb, 2008_clarke_ir_evaluation}.
% }
Therefore, various greedy algorithms are proposed to compute an approximate solution.
For instance,~\citet{2009_yu_swap_approach} proposed the SWAP algorithm considering max-diversification of the items that a recommender system may suggest.
SWAP starts by generating a candidate set and then greedily exchanges items in the candidate set to enhance the diversity score.
Other work selects diverse sets by approximating diversity scores using different strategies~\cite{2003_jain_knn_diversity, 2009_gollapudi_axiomatic_diversity, 2009_leuken_clt}.
% \revision{
Noticeably,~\citet{2011_vieira_divdb} developed GNE and GMC algorithms using a Maximum Marginal Relevance (MMR) scoring function that approximates the max-diversification score to give better diversification results than  prior work. We %test our system against 
use these algorithms in the experiments.
Allan et al.~\cite{2017_borodin_max_sum_diversification} provide theoretical guarantees and approximation bounds for the computation of MMR, which is central to Max-Sum diversification algorithms.
% }
Moreover,~\citet{2015_drosou_disc_diversity} proposed a threshold-based definition of diversity where they consider two items to be similar if they are within a given threshold. They develop an approximate algorithm to select a set of diverse items from a given set of relevant items such that each relevant item is represented in the diverse set by a similar item and each item in the result set is dissimilar to each other. Note that the diverse set could be empty if no items satisfy the given threshold. Hence, we develop a diversity algorithm to output k-diverse items instead of using a threshold-based approach. Note that all of this work considers diversification of image search, web search, and product recommendations. The diversification of tuples, specifically a set of unionable tuples, has not yet been studied. 
These prior IR diversification works assume %smaller -- smaller than what???
relatively small sets of relevant items to diversify and consider effectiveness over efficiency as their use case is to display a small number of diverse search results or recommend a small number of diverse items for human consumption~\cite{2014_vargas_diversity_recommender, 2018_cecchini_topic_retrieval, 2020_gao_toward_fairer_search_results}. 
However, since we may need to yield hundreds of diversified tuples from 
tens of thousands of unionable tuples, we 
% \revision{
consider how to scale diversification algorithms.
% } 

Recently, Large Language Models (LLMs) have shown promising performance in different tabular tasks that need semantic understanding such as column type annotation~\cite{2023_korini_cta_chatgpt}, understanding table unionability~\cite{2023_pal_ugen} and entity matching~\cite{2022_narayan_llm_data_wrangling}.
For tuple diversification, we may prompt an LLM~\cite{2023_franch_ai_vision} with a list of tuples and ask it to return k-diverse tuples. However, as we have to consider thousands of tuples, the current LLMs' input token limits may prohibit their use for this task. Nevertheless, we adopt an LLM baseline~\cite{2020_floridi_gpt3} for an effectiveness experiment in a small dataset. 

\section{Problem Definition and Overview}
\label{section:preliminaries}

% \begin{table}[h]
% \caption{Symbols used in this paper and their definitions}
% \label{tab:notation_table}
% \vspace{-2mm}
% \scalebox{.70}{
% \setlength\tabcolsep{1.5pt}
% \begin{tabular}{cl|cl}
% \hline
% \textbf{Symbol}                    & \textbf{Definition}                           & \textbf{Symbol}                   & \textbf{Definition}                \\ \hline
% $\mathcal{D}$ & Set of Data lake Tables                                 &$q_i$                                & Query Tuple $q_i$                     \\
% $T, T_{i}$                      & Table, the $i_{\text{th}}$ table in $\mathcal{T}$ & $v_i,t_i[c_j]$                  & $i_{th}$ value, value of $t_i$ in Column $c_j$ \\
% $c_j, T.c_j$                        & $j_{th}$ Column, $j_{th}$ Column in table $T$              & $\mathcal T$ ($s \!=\! |\mathcal T|$) & Set of unionable data lake tuples            \\
% $Q, \mathcal Q$                              & Query Table, Set of Query Tuples                        & $\mathcal F$ ($k \!=\! |\mathcal F|$) & Set of k diverse unionable tuples \\
% $t_i$                              & Data lake Tuple $t_i$             & $E(t_i)$                              & Embedding of Tuple $t_i$ \\
% $n$                              & Number of Tuples                   & $E_{\mathcal T}$& List of Embeddings of tuples in $\mathcal{T}$ \\
% \hline
% \end{tabular}}%
% \end{table}

We %represent 
denote a query table using $Q$ and a set of data lake tables using $\mathcal{D}$ where $T \in \mathcal{D}$ represents a data lake table.
A tuple from any data lake table is called a \emph{data lake tuple} ($t$) and a tuple from a query table is called a \emph{query tuple} ($q$).
We use $c$ to represent a column and accordingly, for Table $T$ containing a tuple $t_i$, we use  $T.c_j$ and $t_i[c_j]$ to represent its $j_{th}$ column and the corresponding value of tuple $t_i$ in column $c_j$ respectively.
% \ak{
Furthermore, let $E(t_i)$ be an embedding (representation) of tuple $t_i$ in an embedding space.

\subsection{Tuple Diversity Score}
Let $\delta(.)$ be a tuple distance function that inputs embeddings of two tuples and outputs a distance between them such that the distance between embeddings of a tuple and itself is $0$.
Let, $div(.)$ be a diversity scoring function that maps embeddings of a set of tuples to their diversity score.
We consider $div(.)$ to be the maximum for the optimal set of diverse  tuples. 
% \revision{
In our evaluation, we will use common diversity functions from the literature that include maximizing the embedding distance between all tuples (Max-sum diversification)~\cite{2011_vieira_divdb} or maximizing the minimum distance between tuples (Max-min diversification)~\cite{2021_moumoulidou_diversity_fairness}.
% }

\subsection{Problem Definition}
\label{section:problem_definition}
Given a set of $n$ query tuples, the Diverse Unionable Tuple Search Problem aims to find a set of $k$ unionable tuples in a data lake that are also diverse. 
% \revision{
We adopt the existing unionability definition that 
%Note that we consider 
two tuples are \emph{unionable} if they are either from the same table or two unionable tables, i.e., tables sharing unionable columns~\cite{2018_nargesian_tus, 2023_fan_starmie}.
% }

\begin{definition}[Diverse Unionable Tuple Search Problem]
     Given a Query Table $Q$ having a set of tuples $\{q_1, q_2, \dots q_n \}$, a set of unionable Data Lake Tuples $\{t_1, t_2, \dots t_{s}\}$, a positive integer $k < s$, and a diversity scoring function $div(.)$, the diverse unionable Tuple Search Problem is: find a set of k unionable tuples such that $div(\{q_1, q_2 \dots q_n\} \cup \{t_1, t_2 \dots t_k\})$ is maximized.  
  \end{definition}

\subsection{Solution Overview}
\label{section:solution_overview}

\begin{figure*}
  \includegraphics[scale = 0.5]{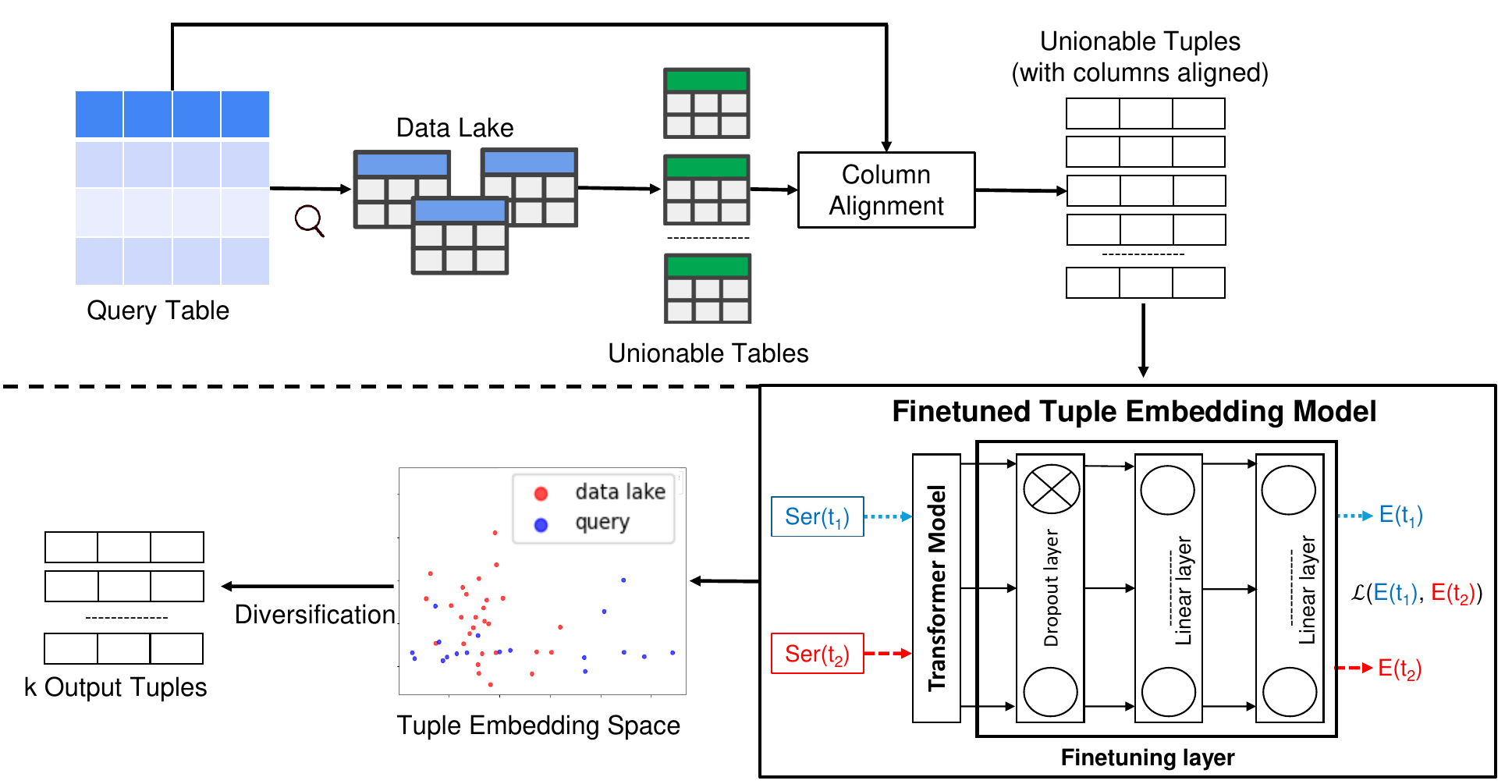}
  \caption{Block diagram of Diverse Unionable Tuple Search (\ourmethod)}
\label{fig:dust_block_diagram}
\end{figure*}

Now, we present a high-level overview of \ourmethod
% ~(\Cref{alg:dust})
, our solution to the Diverse Unionable Tuple Search Problem (shown in~\cref{fig:dust_block_diagram} and ~\cref{alg:dust}).
Given a query table, \ourmethod first obtains a set of unionable tables from the data lake, aligns their columns, and outer-unions the tables. 
A tuple embedding model is applied to %create an 
encode each tuple.
Then, \ourmethod diversifies the tuples to select k-diverse tuples. 
\cref{fig:dust_block_diagram} summarizes
\ourmethod pipeline which consists of three main steps
and we now briefly introduce each step. 

\begin{algorithm}[h]
\caption{\ourmethod}\label{alg:dust}
\setstretch{0.85} % sets line height
\small
\SetAlgoLined
\LinesNumbered
\textbf{Input}: Query Table $Q$, set of Data Lake Tables $\mathcal D$,
% Unionability Parameter $L$, 
Number of output tuples $k$ \\

    \SetKwFunction{Tsearch}{SearchTables}
    \SetKwFunction{Calign}{AlignColumns}
    \SetKwFunction{EmbedT}{EmbedTuples}
    \SetKwFunction{DiverseT}{DiversifyTuples}
    \SetKwProg{Fn}{Function}{:}{}
    \algocomment{Discover data lake tables unionable with the query table.} \\
    $\mathcal{D'} \gets $\Tsearch{$Q$, $\mathcal D$} \\ \label{line:tsearch}
    % , $L$} \\  %\aamod{$U_{Tables}$ or $U_{Tuples}$?}
    \algocomment{Align matching columns of the discovered tables and union them.} \\
    $\mathcal T$ $\gets$ \Calign{$Q$, $\mathcal{D'}$} \\ \label{line:calign}
    \algocomment{Embed each query and data lake tuple.} \\
    $E_{\mathcal Q}, E_{\mathcal{T}} \gets $ \EmbedT{$\mathcal Q$, $\mathcal T$} \\ \label{line:embedt}
    $\mathcal F$ $\gets$ \DiverseT{$E_{Q}, E_{\mathcal T}, k$} \\ \label{line:dtuples}
\end{algorithm}

\introparagraph{Creating Unionable Tuples}
We first explain the top half of~\cref{fig:dust_block_diagram}, where we are provided with a Query Table $Q$.
We search for  unionable tables from the data lake using any table union search technique~\cite{2023_fan_starmie, 2023_khatiwada_santos, 2018_nargesian_tus,2023_hu_autotus}. To union these tables (to form unionable tuples), we perform column alignment. %}
%
%
%\rjm{
As part of the union search, many search algorithms will align each data lake table individually with the query table,
e.g., Starmie uses maximum-weight bipartite matching between each column in a unionable table and the query table~\cite{2023_fan_starmie}. However, most methods do not output column alignments.  Rather than recomputing %these 
pair-wise matches, %, in 
\ourmethod{} %, we 
uses a \emph{holistic column matcher}~\cite{DBLP:reference/bdt/RahmP19} inspired by a recent data integration approach~\cite{2022_khatiwada_alite}.   Holistic matching allows a collective alignment of all the columns in the 
set of unionable tables and the query table.   Unlike in pure integration however~\cite{2022_khatiwada_alite} (where the goal is to integrate a set of tables), in our setting we want to do a targeted alignment to the query table.  We are not interested in columns from data lake tables that align with each other if there is no query column to which they align.   
Our objective is to get a disjoint set of 
%(data lake) 
columns from the unionable tables such that the columns in each set are aligned together and to a single query column.

% \revision{
Similar to Khatiwada et al.~\cite{2022_khatiwada_alite}, we first embed all query and data lake columns using the set of column values. %}  
Over such embeddings, we apply hierarchical clustering to generate a dendrogram that models all possible clusters of columns. 
Note that no two columns from the same table should be aligned together. Therefore, we enforce a constraint during clustering that prohibits clustering columns from the same table together.
Then, an important decision is to select the number of clusters.
%where the maximum is the number of query columns. 
For that, we compute a cluster quality score for each number of clusters and select the one that maximizes the quality.
We measure quality using Silhouette's coefficient~\cite{1987_rousseeuw_silhouttes,2022_khatiwada_alite}.

Note that we only need values from those data lake columns that align with at least one column of the query table. So, after getting 
%all 
the clusters, we discard those that do not contain a column from the query table. This leaves us with a set of clusters, each with a query column and data lake columns that are aligned. If a query column has no match with one or more of the unionable tables,
outer-union will pad the unionable table 
with a null placeholder (e.g., \textit{nan}) to create tuples that can be unioned with the query table.  
Using this alignment, we can outer-union all  unionable tables to form a set of {\bf unionable  tuples}. We provide further details and a block diagram of column alignment phase in \paperorreport{the technical report~\cite{dust_technical_report}}{~\cref{section:appendix_column_alignment_phase}}.
% }.

\begin{example}
\label{example:column_alignment}
    % \revision{    
    Consider Tables in~\cref{fig:park_tables} such that Table (a) is a query table and Tables (b), (c), and (d) reside in the data lake. 
    Suppose we use a union search technique to find the top-2 unionable tables from the data lake. Then Tables (b) and (d) will be returned as the two most unionable tables with Table (a). Next, we input columns of Tables (a), (b), and (d) into the column alignment phase where we get five clusters of columns. Specifically, the first cluster contains three \texttt{Park Name} columns from Tables (a), (b), and (d). The second cluster contains two \texttt{Supervisor} columns from Tables (a) and (b), and \texttt{Supervised By} column from Table (d). The third cluster contains \texttt{City} column from Table (a) and \texttt{Park City} column from Table (d). The fourth cluster contains two \texttt{Country} columns from Tables (a) and (b), and \texttt{Park Country} column from Table (d). We also get a singleton cluster with \texttt{Park Phone} column from Table (d). 
    Since the
    % } 
    % \revision{
    last cluster contains no columns from the query table, it is discarded.
    % } 
    % \revision{
    Then we are left with the first four clusters of columns that are
    % } 
    % \revision{
    assigned with query columns' headers \texttt{Park Name}, \texttt{Supervisor}, \texttt{City}, and \texttt{Country} respectively.
    % } 
\end{example}

\introparagraph{Tuple Representation}
After 
forming unionable tuples, we embed them into a high-dimensional space (bottom right of~\cref{fig:dust_block_diagram}).
We use a novel embedding model (described in \cref{section:tuple_representation})  that maps a tuple to its fixed-dimension embedding. We embed the query table's tuples using the same model and column ordering. 

\introparagraph{Tuple Diversification}
Finally, we use the tuple embeddings to find k-diverse tuples.
Along with %query and data lake tuple 
the embeddings, we are given a diversity scoring function $div(.)$, and positive integer $k$. We input them to a tuple diversification algorithm that return a set of k-diverse data lake tuples. For the most diverse set, the diversity score measured using $div(.)$ is maximized.
% \revision{
In our experiments, we will use existing diversification algorithms from the literature along with a new scalable alternative that we introduce in 
\cref{section:tuple_diversification}.
% }
% \input{sections/tuple-search}
% \input{sections/tuple-representation}
%\section{Unionable Tuple Diversification}
\section{Unionable Tuple Representation}
\label{section:unionable_tuple_diversification}

%\subsection{Tuple Representation}
\label{section:tuple_representation}

While diversification is a well-studied topic, diversifying unionable tuples is not.  To apply the solution overviewed in the last section, we need to be able to compute a meaningful distance measure that models diversity
(more distant means more diverse). 
Thus, we now create an embedding space to represent tuples, allowing the computation of distances.  
The diversity literature~\cite{2008_xu_diversity_user_study, 2011_vieira_divdb} considers relevance and diversity as opposite dimensions, i.e., increasing relevance is considered to reduce diversity and vice-versa. Using this lens, two similar unionable tuples should be closer to each other in the embedding space and less diverse than two different unionable tuples.
For example, the tuple referring to River Park in~\cref{fig:park_tables} (a) should be much closer to the tuple referring to River Park in~\cref{fig:park_tables} (b) than the tuple referring to Chippewa Park in~\cref{fig:park_tables} (d). 
Now, since in this step we already assume the tuples are unionable, our goal would be to utilize this embedding space to position more distant, yet unionable tuples on the top of the diversity list. For example, given the aforementioned River Park tuple in~\cref{fig:park_tables} (a), we will use the embedding space to consider Chippewa Park in~\cref{fig:park_tables} (d) more diverse from the query table than the River Park tuple in~\cref{fig:park_tables} (b). In this case, both tuples are unionable, but as the latter is identical to a query tuple, the former should be preferred, being both unionable and more diverse. 

To represent a tuple, it is important to consider what columns it contains and the context in which it appears. In related tasks such as entity matching~\cite{2020_li_ditto}, column annotation~\cite{2021_suhara_doduo} and even table union search~\cite{2023_fan_starmie}, pre-trained language models (e.g., BERT~\cite{2019_devlin_bert}) have been shown to correctly capture context. 
It would be reasonable to use them to capture an intuitive notion of similarity that comes from natural language (tuples using similar words will be closer in the embedding space). 
If we want to use an embedding model to embed unionable tuples for diversification, the model must understand if two tuples are unionable so that it can encode their distances accordingly.
As our experiments indicate (Section~\ref{section:experiments}), without proper fine-tuning, existing embedding models do not work well for capturing unionability.
In what follows, we formulate a new fine-tuning technique aiming at classifying unionable tuples. 
Such a fine-tuned model could provide better performance by measuring how similar or diverse unionable tuples are. In addition to using an annotated table union search benchmark~\cite{2018_nargesian_tus}, we also use, in a self-supervision fashion, the fact the tuples originating from the same table are unionable. 
To that end, we create fine-tuned tuple embeddings, described next.

%\subsection{Dataset Preparation}
\paragraph{Dataset Preparation}  We used a table union search benchmark~\cite{2018_nargesian_tus} to create our training data.
The benchmark contains labels indicating if two tables are unionable.
Based on our classification task, we devise a fine-tuning dataset where each data point contains a pair of tuples and a binary label (0 if they originate from non-unionable tables or 1 if they are from the same table or a pair of unionable tables). 
So for comprehensive fine-tuning, we create 
data points of similar tuples by selecting a pair of tuples from the same table or from two unionable tables. 
Furthermore, two tuples from a pair of non-unionable tables in the benchmark are about different topics and hence, diverse from each other. So, for the diverse tuple data points, we create tuple pairs by selecting a tuple each from two non-unionable tables.
The data points are then divided into train, test, and validation sets without leakage.
%\subsection{Model Architecture}
% \begin{figure}[t]
%     \centering
% \includegraphics[scale=.55]{figures/finetuning_architecture.pdf}
%     \caption{Architecture of \ourmethod model for representing unionable tuples}
%     \label{fig:finetuning_architecture}
% \end{figure}

%\subsection{Serialization}
\paragraph{Serialization}
Here, we describe how we input the tuples to the pre-trained model for fine-tuning.
Pre-trained language models are built over textual data and they take natural language sentences as input. In comparison to text, tabular data has different semantics and information encoding principles~\cite{2023_fan_starmie, 2021_suhara_doduo, 2023_dong_deepjoin}.
So, we serialize tuples into sentences to input them in the pre-trained model by retaining tabular properties. Specifically, two tuples may not have the same columns but may still be unioned on a subset of their columns. Hence, to help the model learn unionability on different numbers of columns, we serialize each tuple by separating each column and its value.\footnote{We have experimented with other serializations offline, which were less effective.}
Precisely, let us consider a tuple $t$  having $n$ columns. Let $c_i$ and $v_i$ represent the column's header and value respectively of column $i$~\cite{2019_devlin_bert}. Also, [CLS] and [SEP] are the special BERT-based model tokens that represent the start of a sequence and a separation between the tokens in the sequence respectively. 
We feed the serialized tuple $t$ to the input layer of the BERT-based model as:
$Ser(t)$ :- [CLS] $c_1$ $v_1$ [SEP] $c_2$ $v_2$ $\dots$ [SEP] $c_n$ $v_n$ [SEP]

\begin{example}
\label{example:serialization}
    Consider Tuples in Tables of~\cref{fig:park_tables}. Recall from Example~\ref{example:column_alignment} that we assigned 
    %\texttt{ColA}, \texttt{ColB}, \texttt{ColC} and \texttt{ColD} as dummy 
    \texttt{Park Name}, \texttt{Supervisor}, \texttt{City}, and \texttt{Country} as
    column headers to the columns aligned with respective Query Columns (Table (a)'s columns).
    %: \texttt{Park Name}, \texttt{Supervisor}, \texttt{City} and \texttt{Country} respectively. 
    So, we serialize Tuple (River Park, Vera Onate, Fresno, USA) in Table (a) as: [CLS] Park Name River Park [SEP] Supervisor Vera Onate [SEP] City Fresno [SEP] Country USA [SEP]. Moreover, for Tuple (Chippewa Park, "Brandon, MN", USA, 773 731-0380, Tim Erickson) in Table (d), we use those columns (and order) for serialization that aligned with the Query Table, i.e., all but \texttt{Park Phone}.
    %\texttt{Park Name} and \texttt{Park City}. 
    So, it is serialized as: [CLS] Park Name Chippewa Park [SEP] City Brandon, MN [SEP] Country USA [SEP].
\end{example}

%\subsubsection{Fine-tuning Architecture}
\paragraph{Fine-tuning Architecture}
\label{section:model_architecture}
We want to create a Tuple Embedding Space such that a pair of diverse tuples are farther from each other than a pair of similar tuples.
Our empirical evaluation shows that the pre-trained models as-is are not good at this task (see~\cref{section:tuple_representation_evaluation}).
Therefore, as shown in~\cref{fig:dust_block_diagram} (bottom-right corner), we append a fine-tuning layer to the pre-trained transformer model and train it to embed the tuples such that the diverse tuples are far from each other and vice-versa. For fine-tuning, we test combinations of different layers and parameters empirically and use the best architecture.
% that best represents tuple unionability. 
%aamod: space saving
Specifically, in the \ourmethod model, we append a dropout layer to the regular transformer model which possibly helps us avoid overfitting during training. Then, we pass the dropout output through two linear layers such that the outcome of the last linear layer is a fixed-dimension embedding of the tuple. 

% \revision{
Note that we use the fine-tuned model to embed each tuple individually for diversification in the later phase~(\Cref{section:tuple_diversification}). So, during the training process, given a data point with a pair of tuples and a binary label, we pass each serialized tuple one after another through the model to represent them independently. As shown in~\cref{fig:dust_block_diagram} (bottom right) for tuples $t_1$ and $t_2$, we pass $Ser(t_1)$ and $Ser(t_2)$ one after another to get their representations $E(t_1)$ and $E(t_2)$ respectively. 
Then, we compute loss by comparing the distance between two representations against the ground truth labels. The loss is fed back to update the model parameters. 
% }
Our architecture is flexible to use any distance function. For our experiments, we
use cosine distance, and accordingly, measure cosine embedding loss, 
$\mathcal{L}(E(t_1), E(t_2)) = \begin{cases}
1 - \cos(E(t_1), E(t_2)) & \text{if label } = 1 \\
\max(0, \cos(E(t_1), E(t_2))) & \text{if label } = 0
\end{cases}$, where $\cos(E(t_1), E(t_2)) = \frac{E(t_1) \cdot E(t_2)}{\|E(t_1)\| \|E(t_2)\|}$.
\footnote{In our experiments, we use PyTorch's implementation (\url{https://pytorch.org/docs/stable/generated/torch.nn.CosineEmbeddingLoss.html}).}

% \begin{equation}
%     \mathcal{L}(E(t_1), E(t_2)) = \begin{cases}
%       1 - \cos\{E(t_1), E(t_2)\} & \text{if label = 1}\\
%       \max[0, \cos\{E(t_1), E(t_2)\}] & \text{if label = 0}\\
%     \end{cases} 
% \end{equation}
% where, $\cos[E(t_1), E(t_2)] = \frac{E(t_1) \cdot E(t_2)}{\|E(t_1)\| \|E(t_2)\|}$

\section{Scalable Tuple Diversification}
\label{section:tuple_diversification}
% \revision{
As we discussed, % in~\cref{section:solution_overview},
tuple diversification is computationally hard~\cite{2011_vieira_divdb} 
% \revision{
and has traditionally been applied in IR to smallish sets (which have less than a few hundred of elements).
% \revision{
Because we wish to consider in our experiments larger sets (with thousands of elements), here
% }
 we propose a clustering-based approach that efficiently computes an approximate solution. 
We also propose a pruning method that controls the number of candidate diverse tuples input to the clustering algorithm.
% }
Our algorithm~(\cref{alg:diversification})
takes as input
two separate sets of embeddings for the unionable data lake tuples and 
%unionable 
query tuples, along with a positive integer k, and it returns k-diverse unionable data lake tuples.
\begin{algorithm}[h]
\caption{DiversifyTuples}\label{alg:diversification}
\setstretch{0.85} % sets line height
\small
\SetAlgoLined
\LinesNumbered
\textbf{Input}: set of query tuples embeddings $E_{\mathcal Q}$, set of unionable data lake tuples embeddings $E_{\mathcal T}$, positive integer $p$, Number of output tuples $k$, Number of unionable tuples $s$ \\
    \SetKwFunction{ptuples}{PruneTuples}
    \SetKwFunction{ctuples}{ClusterTuples}
    \SetKwFunction{mdistance}{PairwiseDistance}
    \SetKwFunction{topk}{FindTopK}
    \SetKwProg{Fn}{Function}{:}{}
    % \algocomment{Apply standard deviation.} \\
    $E_{\mathcal{T'}} \gets$ \ptuples{$E_{\mathcal T}$, $s$} \label{line:div_ptuples}\\
    \algocomment{Cluster data lake tuples and select candidates.} \\
    $E_{\mathcal{T'}} \gets$ \ctuples{$E_{\mathcal T'}$, $k \cdot p$} \label{line:div_ctuples}\\
    \algocomment{Compute ranking score for each candidate data lake tuples}\\
    $rank\_scores \gets () $\\ \label{line:start_rerank}
    \For {$t \in \mathcal{T'}$} {
        $rank\_scores[t].insert(0)$
    }
    \For {$t \in \mathcal{T'}$} 
    {
        \For{$q \in \mathcal{Q}$} 
        {
            $rank\_scores[t].update(\min(rank\_scores[t], \delta(t, q)))$
        }
    } \label{line:end_rerank}
    
    %$pairwise\_distance \gets $ \mdistance{$E_{\mathcal Q}, E_{\mathcal{T'}}$} \label{line:div_pdistance}\\ 
    \algocomment{Return top-k data lake tuples}\\
    $\mathcal{F} \gets$ \topk{$rank\_scores$, $k$} \label{line:div_rerank} \\
\end{algorithm}
% \revision{
Recall that we want those data lake tuples that provide potentially new information to the query table. We consider two things when selecting the diverse tuples. First, we want to select
a set of
unionable data lake tuples such that the selected tuples are diverse among themselves. This helps us to ensure that the newly added tuples do not bring redundancy among themselves. Second, we ensure that the selected tuples
are diverse from the set of tuples
present in the query table. 
% }
\cref{alg:diversification} first prunes the set of unionable tuples (\cref{line:div_ptuples} and \cref{sec:prun}).  
Then, we compute candidate
%shortlist data lake 
unionable tuples that are diverse among themselves using clustering(\cref{line:div_ctuples} and \cref{sec:candidates}). We compare these candidates against the query tuples and re-rank them such that the selected unionable tuples
% adding information unavailable in the
are diverse from the
query tuples~(\cref{line:div_rerank} and \cref{sec:rank}).

\subsection{Pruning Candidate Data Lake Tuples}\label{sec:prun}
 Each unionable data lake table can have a large number of tuples and the step of clustering them within our diversification algorithm can be time-consuming. So, we begin by pruning and restricting the number of tuples before initiating clustering~(\cref{line:div_ptuples}). Note that we want to output data lake tuples that are diverse from each other. Specifically, when a tuple's embedding is significantly distant from others in the table, it signifies greater diversity than its counterparts.
We initially compute the mean embedding of all eligible data lake tuples within each table for diversification. Then, for every tuple, we compute its distance from this mean embedding and rank them accordingly. Mathematically, for a tuple $t$ from Table $T$ having $E(t_m)$ as the mean embedding of its tuples, the rank score of $t$ is computed as $Score(t) = \delta(E(t_m), E(t))$.
% \begin{equation}
%     Score(t) = \delta(E(t_m), E(t))
% \end{equation}
% where, $E(t_m)$ is the mean embedding of tuples in Table $T$.
% computed as $\frac{1}{|T|} \sum_{i = 1}^{|T|} E(t_i)$.
The top-$s$ tuples based on this ranking are then selected for clustering. As our pruning method prioritizes tuples with the highest distances from each other, it ensures the retention of the most diverse candidates for further processing.

\subsection{Clustering Candidate Data Lake Tuples}\label{sec:candidates}

To select a set of diverse candidate unionable tuples, we want to select  candidates that are far from each other in the embedding space. For this, we apply clustering over the set of unionable data lake tuples. Precisely, we use hierarchical clustering as it can scale well for a reasonable number of clusters (in our case the number of output diverse tuples, k).
To ensure that we select enough candidates, we use a parameter $p$ to control the number of clusters 
%thus and so 
such that 
there are more than k candidate tuples.
In~\cref{alg:diversification}, 
the number of clusters (size of candidate tuple set)
$\mathcal{T'} = k \cdot p$ (see~\cref{line:div_ctuples}). As each cluster contains similar tuples and the tuples in different clusters are diverse, we select a representative diverse tuple from each cluster. Specifically, to increase the distance between candidate tuples, we select each cluster's medoid, which is the central-most element of the cluster. The medoids then form a candidate tuple set containing data lake tuples that are diverse among themselves. 
% \revision{
Selecting the medoid of each cluster as a candidate diverse tuple makes the approach more robust to outliers and this could also be augmented with an outlier detection phase before the clustering~\cite{2020_smiti_outlier_detection}.
% }

\subsection{Re-ranking Candidate Diverse Tuples}\label{sec:rank}

Among the candidate data lake tuples that are diverse among themselves, now we want to select tuples that are 
%least redundant 
most diverse 
with respect to the query tuples.
\ak{
A candidate unionable data lake tuple can be diverse with one query tuple but not with other query tuple. Hence, we want to ensure that a selected unionable tuple is not redundant with any query tuples.   For each candidate unionable tuple, we compute its distance from each query tuple and then assign a score which is its minimum distance from all the query tuples~(\cref{line:start_rerank}-\ref{line:end_rerank} of \cref{alg:diversification}). }
We then sort the candidate tuples in descending order such that the best-ranked tuple has the maximum minimal distance from the query tuples. 
In case of a tie, we prioritize the tuple with the highest average distance from all query tuples.
Finally, the set of top-ranked k unionable tuples in the ranking is returned as output~(\cref{line:div_rerank}).

\begin{figure}[h]
  \includegraphics[scale = 0.3]{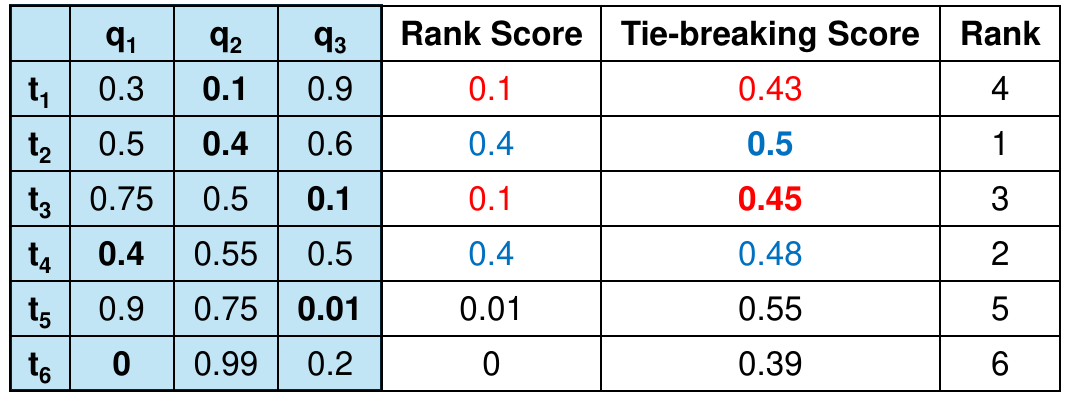}
  \caption{Ranking candidate diverse unionable tuples.}
\label{fig:diversification_example}
\end{figure}

\begin{example}
\label{example:diversification_example}
Consider query tuples $q_1$, $q_2$ and $q_3$ and %6 candidate 
data lake tuples $t_1$, $t_2 \dots t_6$.~\cref{fig:diversification_example} shows the tuple distance between each query and data lake tuple, and we want to rank the candidate tuples.
First, we compute the \texttt{Rank Score} for each candidate tuple which is the minimum of its distance to all the query tuples. 
We also compute the average distance between each candidate tuple and query tuples to break ties (\texttt{Tie-breaking Score}). 
Tuples $t_2$ and $t_4$ have the highest \texttt{Rank Score} i.e. $0.4$. Since both tuples have equal score, we look at Tie-breaking Score and rank $t_2$ (score of $0.5$) above $t_4$ (score of $0.48$). Similarly, Tuple $t_3$ is ranked before Tuple $t_4$. Tuples $t_5$ and $t_6$ with the least Rank Scores ($0.01$ and $0$) are ranked at the bottom.
\end{example}

\subsection{Tuple Diversification Evaluation}\label{sec:div_eval}

In the literature, the objective of diversification is to simply maximize diversity among all results.  Hence, algorithms are evaluated mainly on two criteria: Max-sum diversification where the sum of distances between items (tuples) in the diverse set is maximized~\cite{2017_borodin_max_sum_diversification}, and Max-min diversification where the minimum distance between the diverse items in the set is maximized~\cite{2021_moumoulidou_diversity_fairness, 2019_yang_diveristy_constraints}. 
However, for diverse unionable tuple search, we have a different objective:  select unionable tuples that are \emph{as diverse as possible} with query tuples and \emph{as diverse as possible} to each other.
So, based on those criteria, we present two adapted metrics to evaluate tuple diversification. 

\textbf{(i) Average Diversity.} To evaluate if the distance between  tuples is maximized (Max-sum diversification criteria), we compute the average distance between tuples in a diverse set. 
% \revision{
Precisely, we compute the average distance between query table tuples and the data lake table tuples, and between data lake table tuples.
% } 
An optimal algorithm maximizes the distance between the tuples and hence gives the highest average diversity score.
    Given a set of Query Tuples $\{q_1, q_2 \dots q_n\}$ and a set of k-diverse unionable data lake tuples $\{t_1, t_2 \dots t_k\}$ returned by a method, then the Average Diversity is computed as:
    % $\textrm{Average Diversity} = \frac{\sum_{i = 1}^n \sum_{j = 1, i \neq j}^k \delta(q_i, t_j) + \sum_{i = 1}^{k-1} \sum_{j = i+1}^k \delta(t_i, t_j)}{n + k}$.
    % aamod: space
    \begin{equation}
        \textrm{Average Diversity} = \scriptstyle\frac{\sum_{i = 1}^n \sum_{j = 1, i \neq j}^k \delta(q_i, t_j) + \sum_{i = 1}^{k-1} \sum_{j = i+1}^k \delta(t_i, t_j)}{n + k}
        \label{eq:average_diversity}
    \end{equation}
    % % \revision{
    As query tuples are given %by the users 
    and the distance between them is constant for all algorithms, we exclude their distance from the evaluation. 
    % }

\textbf{(ii) Min Diversity.} Based on the Max-min Diversification criteria, i.e., maximizing the minimum distance between the items in the diverse set, we also compare the minimum distance among the tuples selected in a k-diverse set. An optimal algorithm gives the highest Min Diversity score.
Mathematically, for query tuples $\{q_1, q_2 \dots q_n\}$ and a set of k-diverse unionable data lake tuples $\{t_1, t_2 \dots t_k\}$ returned by a method,
then $S_{Q} = \{\delta (q_i , t_j) \mid i \in [1, n], j \in [1 , k] \} $, $S_{T} = \{\delta (t_i , t_j) \mid i \in [1, k-1], j \in [i+1 , k]\}$,
 % and $ \textrm{Min Diversity} = \min \{S_Q \cup S_T \}$.
% \vspace{-.125in}
% \begin{equation*}
%         S_{Q} = \{\delta (q_i , t_j) \mid i \in [1, n], j \in [1 , k] \}    
% \end{equation*}
% \vspace{-.125in}
% \begin{equation*}
%         S_{T} = \{\delta (t_i , t_j) \mid i \in [1, k-1], j \in [i+1 , k]\} \\     
% \end{equation*}
% \vspace{-.1in}
\begin{equation}
        \textrm{Min Diversity} = \min \{S_Q \cup S_T \}     
\end{equation}

Prior work~\cite{2011_vieira_divdb} also considers the trade-off between relevance and diversity controlled by a user-defined parameter. Accordingly, they formulate a diversity scoring function that inputs a set of k-diverse items and returns their diversity score. The optimal set of k-diverse items gives the maximum diversity score. As determining such an optimal set is computationally hard, they created ground truth considering $k = 5$, by using a brute force approach over 5 different queries each having 200 candidate relevant items. They report precision and the gap between the diversification scores of the diverse set returned by an algorithm and the ground truth set. In our case, we are looking to rank thousands of tuples and thus such an evaluation framework is not feasible.
\section{Experiments}
\label{section:experiments}

We now empirically evaluate \ourmethod{} against different baselines.\footnote{\label{footnote:dust_github}\url{https://github.com/northeastern-datalab/dust}}
We run all our experiments using Python 3.8 on a server with
Intel(R) Xeon(R) Gold 5218 CPU @ 2.30GHz
processor and $4 \times 8$ GB Tesla GPUs. We start by describing the benchmarks that we use for our evaluation (\cref{section:main_benchmarks}). Next, we investigate 
%embedding models for 
column alignment (\cref{section:column_alignment_evaluation}) using several different existing column embeddings 
%for this task 
%aamod: space saving
and comparing our holistic approach to a state-of-the-art pairwise matcher from Starmie~\cite{2023_fan_starmie}. %Importantly, 
Then, we %intend 
aim to answer the following: 
\textbf{RQ1.} As a sanity check, we first consider how effective is \ourmethod in distinguishing  unionable tuples against  existing tuple embedding techniques~(\cref{section:tuple_representation_evaluation})?
\textbf{RQ2.} How well do existing diversification algorithms perform when diversifying unionable tuples?  Do the new more scalable \ourmethod diversification algorithm achieve the goal of scaling diversification to larger numbers of tuples and if so, what is the impact on effectivness 
(\cref{section:tuple_diversification_evaluation})?
\textbf{RQ3.} How effective and efficient is \ourmethod in the end-to-end task of finding  diverse unionable tuples in comparison with existing unionable table search techniques~(\cref{section:tuple_search_evaluations})?
\textbf{RQ4.} Does \ourmethod provide practical and intuitive benefit over existing unionable table search techniques (\cref{section:case_study})?
% \begin{enumerate}[label=RQ\arabic*.]
%     \item As a sanity check, we first consider how effective is \ourmethod in distinguishing  unionable tuples against  existing tuple embedding techniques~(\cref{section:tuple_representation_evaluation})?
%     \item How well do existing diversification algorithms perform when diversifying unionable tuples?  Do the new more scalable \ourmethod diversification algorithm achieve the goal of scaling diversification to larger numbers of tuples and if so, what is the impact on effectivness 
% (\cref{section:tuple_diversification_evaluation})?
%     \item How effective and efficient is \ourmethod in the end-to-end task of finding  diverse unionable tuples in comparison with existing unionable table search techniques~(\cref{section:tuple_search_evaluations})?
%     \item Does \ourmethod provide practical and intuitive benefit over existing unionable table search techniques (\cref{section:case_study})?
%     % \item How efficient is \ourmethod in diversifying the tuples against the existing diversifying techniques?
% \end{enumerate}

As each experimental setup and baselines differ across research questions, they are discussed within the corresponding subsections. 

\subsection{Benchmarks}
\label{section:main_benchmarks}

% \begin{table*}
% \small
% % \setlength{\tabcolsep}{1pt}
% \caption{Benchmarks used in the experiments. 
% }
% \begin{tabular}{l|ccc|ccc|c}
% \toprule
%     \multirow{2}{*}{\makecell[l]{\textbf{Benchmark}}}&
%     \multicolumn{3}{ c| }{\textbf{Query}}&
%     \multicolumn{3}{ c| }{\textbf{Data Lake}}&
%     \multicolumn{1}{ c }{\textbf{\# Average Unionable}}\\
%     &\textbf{\#Tables}&\textbf{\#Columns}&\textbf{\#Tuples}&\textbf{\#Tables}&\textbf{\#Columns}&\textbf{\#Tuples}&\textbf{Tables Per Query}\\
% \hline
% %TUS (Small)&125&1.6 K&557 K&1530&14.8 K&6.8 M& 188\\
% TUS&125&1.6K&557K&5044&55.5K&9.6M&188\\
% TUS-Sampled&30&355&134K&233&3.1K&1M&10\\
% SANTOS&50&615&1.07M&550&6.3K&3.8M&14\\
% UGEN-V1&50&400&550&1000&8K&10K&10\\ 
% \bottomrule
% \end{tabular}
% \label{tab:main_benchmarks}
% \end{table*}

\begin{figure}
  \includegraphics[scale = 0.45]{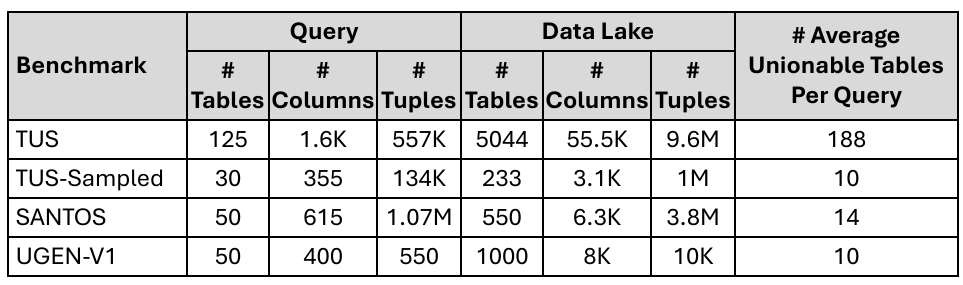}
  \caption{Benchmarks used in the experiments.}
\label{fig:main_benchmarks}
\end{figure}
We experiment using table union search benchmarks from the literature~\cite{2018_nargesian_tus, 2023_khatiwada_santos, 2023_pal_ugen}, each %. Each benchmark 
comes with a set of query tables and a labeled set of unionable data lake tables. \cref{fig:main_benchmarks} details the number of tables, columns, and tuples in each benchmark.

\subsubsection{\textbf{TUS Benchmark}~\cite{2018_nargesian_tus}} 
The TUS benchmark contains over %5 thousand 
5,000 tables generated by selecting and projecting rows and columns from 32 non-unionable base tables.  The generated tables originating  from the same base table are unionable, those from different base tables are non-unionable.\footnote{\url{https://github.com/RJMillerLab/table-union-search-benchmark}} We create two variations of the TUS Benchmark based on our experimental requirements.

\introparagraph{TUS-Sampled Benchmark}
On average, the TUS Benchmark has over 188 unionable data lake tables per query table, significantly higher than 
other benchmarks.
So to test effectiveness of some non-scalable methods, we created {\em TUS-Sampled}
%Therefore, we create a smaller variation 
by selecting 30 query tables and sampling 10 unionable tables per query table.

\introparagraph{TUS Fine-tuning Benchmark}
To build the \ourmethod tuple representation model (see~\cref{section:tuple_representation}), we create a fine-tuning benchmark using tables and unionability ground truth from TUS benchmark~\cite{2018_nargesian_tus}. The created benchmark consists of 60K data points where each data point consists of a tuple pair and a unionability label (0 or 1) representing whether the tuples are unionable. The benchmark is balanced, i.e., it contains 30K unionable and non-unionable tuple pairs each. The tuple pairs selected from the same table or two unionable tables are considered unionable and the tuples from two non-unionable tables are considered non-unionable. We divide data points into train/test/validation sets in a popularly used ratio of 70:15:15 (42K, 9K, and 9K data points respectively). We ensure all three sets are balanced regarding the number of unionable and non-unionable tuple pairs. We also ensure that there is no data leakage between train, test, and validation sets.

\subsubsection{\textbf{SANTOS Benchmark}~\cite{2023_khatiwada_santos}}
This benchmark contains 550 data lake tables and 50 query tables created by projecting and selecting rows and columns of 297 base tables from Canada, US, UK, and Australian Open Data.\footnote{\url{https://zenodo.org/records/7758091}}
The tables are created following the TUS benchmark creation approach. 
But 
unlike
TUS, when creating the tables, SANTOS also considers the binary relationships between the column pairs such that the unionable tables not only contain unionable columns, but also share at least one binary relationship.

\subsubsection{\textbf{UGEN-V1 Benchmark}~\cite{2023_pal_ugen}}
The UGEN-V1 benchmark is a table union search benchmark that is generated using a Large Language Model~\cite{2020_floridi_gpt3}. 
It contains 50 query tables generated from different topics and each table has 10 unionable data lake tables and 10 non-unionable tables on the same topic. As LLMs are not always accurate, the authors also provide a manually verified ground truth which we use in our experiments.\footnote{\url{https://github.com/northeastern-datalab/gen}} 

Finally, we remove the columns having all null values and query tables that have less than 3 rows from our experiments.

\subsection{Column Alignment Evaluation}
\label{section:column_alignment_evaluation}

\begin{table*}
% \small
% \fontsize{8}{10}\selectfont
% \setlength{\tabcolsep}{1pt}
\centering
\caption{Column Alignment effectiveness. %of different models in different benchmarks. 
%The b
Best score along each column is in bold; the second best score is \underline{underlined}.}
  \label{tab:column_alignment_results}
  \begin{tabular}{l|l|rr>{\columncolor[gray]{0.8}}r|rr>{\columncolor[gray]{0.8}}r|rr>{\columncolor[gray]{0.8}}r}
    \toprule
    \multirow{2}{*}{\makecell[l]{\textbf{Serialization}}}&\multirow{2}{*}{\makecell[l]{\textbf{Model}}}&\multicolumn{3}{ c| }{\textbf{TUS-Sampled}}& \multicolumn{3}{ c| }{\textbf{SANTOS}}& \multicolumn{3}{ c }{\textbf{UGEN-V1}}\\
    &&\multicolumn{1}{c}{\textbf{P}}&\multicolumn{1}{c}{\textbf{R}}&\multicolumn{1}{c|}{\textbf{F1}}&\multicolumn{1}{c}{\textbf{P}}&\multicolumn{1}{c}{\textbf{R}}&\multicolumn{1}{c|}{\textbf{F1}}&\multicolumn{1}{c}{\textbf{P}}&\multicolumn{1}{c}{\textbf{R}}&\multicolumn{1}{c}{\textbf{F1}}\\
    \hline
    {\multirow{5}{*}{\makecell[l]{\textbf{Cell-level}}}}&FastText&0.86&0.60&0.66&0.64&0.86&0.70&0.36&0.78&0.43\\
    &Glove&0.59&0.83&0.63&0.65&0.84&0.71&0.40&0.76&0.43\\
    &BERT&0.60&0.60&0.59&0.57&0.68&0.60&0.41&0.61&0.44\\
    &RoBERTa&0.61&0.80&0.69&0.57&0.84&0.66&0.52&0.68&0.53\\
    &sBERT&0.67&0.77&\underline{0.70}&0.60&0.86&0.69&0.50&0.71&0.52\\
    \hline
    {\multirow{3}{*}{\makecell[l]{\textbf{Column-level}}}}
    &BERT&0.80&0.57&0.64&0.68&0.68&0.66&0.49&0.57&0.47\\
    &RoBERTa&0.81&0.72&\textbf{0.74}&0.71&0.87&\textbf{0.76}&0.58&0.66&\textbf{0.58}\\
    &sBERT&0.74&0.72&0.68&0.71&0.86&\underline{0.76}&0.54&0.71&\underline{0.58}\\
    \hline
    {\multirow{2}{*}{\makecell[l]{\textbf{Table context}}}}
    &Starmie (B)&0.30&0.67&0.41&0.23&0.53&0.32&0.15&0.76&0.24\\
    &Starmie (H)&0.83&0.43&0.55&0.43&0.33&0.18&0.64&0.62&0.57\\
    %\cline{2-11}
    
  \bottomrule
\end{tabular}
\end{table*}

For column alignment, we use existing embedding models to embed the columns. So, we empirically evaluate different embeddings for aligning unionable data lake columns with the query table columns.

\subsubsection{Experimental Setup}
Recall that to align columns, we first represent each column in an embedding space using pre-trained embedding models. Then, based on their embeddings, we cluster them into disjoint set of columns. So, we evaluate the performance of different embedding methods when they are used to represent the columns and align them.
For clustering, we use the Agglomerative Clustering module available in Scikit-learn's library.\footnote{\url{https://scikit-learn.org/stable/modules/classes.html\#module-sklearn.cluster}}
We report results using average linkage and Euclidean distance based on empirical effectiveness. 
Following the literature~\cite{2022_khatiwada_alite}, we use Silhouette's Coefficient to measure the cluster quality~\cite{1987_rousseeuw_silhouttes}.

\subsubsection{Evaluation Metrics} 
Similar to the prior work~\cite{2022_khatiwada_alite}, we report Precision (P), Recall (R), and F1-Score (F1) using different embeddings for column alignment. Recall that we discard clusters without a query column (see~\cref{section:preliminaries}). Accordingly, we want to evaluate if 
%the columns in 
the data lake columns are assigned to clusters with their correct aligning query column. The ground truth contains all true column alignments in the form of column pairs. This includes column pairs formed by each query column with its aligning data lake columns, and column pairs formed by two data lake columns that have the same matching query column. Furthermore, it is important to distinguish the query columns having no matching data lake columns. So, we also include each query column with no match in the ground truth. In the same way, we form a set of column alignments using the clusters produced by a method.
Let, $A_G$ be the set of true column alignments in the ground truth. 
Let, $A_M$ be the set of column alignments given by a method. Then, we compute \textbf{P}, \textbf{R} and $\mathbf{F_1}$ as:
$P = \frac{{A}_{G}\cap{A}_{M}}{{A}_{M}}, R = \frac{{A}_{G}\cap{A}_{M}}{{A}_{G}}, F_1 = \frac{2 \cdot P \cdot R}{P + R}$.

% \begin{equation}\label{eq:P_R_column_alignment}
%     P = \frac{{A}_{G}\cap{A}_{M}}{{A}_{M}}, R = \frac{{A}_{G}\cap{A}_{M}}{{A}_{G}}, F_1 = \frac{2 \cdot P \cdot R}{P + R}
% \end{equation}

\subsubsection{Baselines} 
Now, we describe different embedding models that we test for the column alignment task. We use two word embedding models: \textbf{FastText}~\cite{2016_joulin_fasttext} and \textbf{Glove}~\cite{2014_pennington_glove} for our experiments. Furthermore, we also experiment using three language models: \textbf{BERT}~\cite{2019_devlin_bert}, \textbf{RoBERTa}~\cite{2019_liu_roberta} and Sentence BERT (\textbf{sBERT})~\cite{2019_reimers_sbert}. For all models, we create a \textbf{Cell-level} variation where we embed a column by first representing each cell value independently and then averaging the representation of all the values within a column. 
Furthermore, for language model baselines, we also create a \textbf{Column-level} variation. Specifically, rather than treating each cell independently, the \textbf{Column-level} variation concatenates all cell values into a sentence and inputs it into the model. 
The model then returns a representation for the column. Note that all three language models have a token limit of 512. 
Therefore, using all the column values in the input may not be possible. Hence, we follow the literature~\cite{2023_fan_starmie, 2023_dong_deepjoin, 2021_suhara_doduo} and select at most 512 most representative tokens for each column based on their TF-IDF scores~\cite{2003_ramos_tfidf}.
Moreover, table union search techniques such as Starmie~\cite{2023_fan_starmie}, assess table unionability based on the maximum-weight bipartite matching between query and candidate data lake tables' columns. 
Starmie embeds each column to capture the entire \textbf{Table context}.
So, we use Starmie as a baseline, implementing two versions: \textbf{Starmie (B)}, by applying \textit{bipartite} matching between the columns embedded using Starmie. Next, we also implement \textbf{Starmie (H)} that uses our \textit{holistic} column alignment approach on Starmie's column embeddings.

\subsubsection{Effectiveness}
We now evaluate the effectiveness of different embedding methods on the column alignment task. Specifically, we report precision, recall, and F1-Score using each embedding method in~\cref{tab:column_alignment_results}. For our discussion, we focus on the combined metric i.e., F1-Score (highlighted in gray). We observe that the Column-level RoBERTa performs the best in aligning the columns as it has the highest F1-score in all the benchmarks. Specifically, RoBERTa outperforms the second-best method, the Column-level variation of sBERT, by around 8\% in TUS-Sampled. In SANTOS and UGEN-V1 benchmarks, sBERT also performs as well as RoBERTa. BERT, possibly because it is the smallest among the three language models, has the lowest F1-score among them.
Further, we observe that for all three language models, the Column-level variation is more effective than the Cell-level variation in terms of F1-score in almost all cases; the only exception is sBERT in TUS-Sampled by a small margin. 
\ak{
A possible reason for Column-level variation's better performance 
is that the Column-level variation gets more tokens as input at once, and accordingly, it becomes easier for them to understand the column than for Cell-level variation where they receive tokens only from one cell at a time to understand the context.}
Moreover, the word embedding baselines (FastText and Glove) give comparative performance to cell-level language models in all the benchmarks in terms of F1-score, but they are comprehensively outperformed by column-level RoBERTa and column-level sBERT. This also shows that larger language models 
%when receive enough tokens, 
can contextualize the columns better to align them. 
Additionally, both variations of Starmie are mostly outperformed by other models. This may be attributed to Starmie embedding each column with the context of the entire table, resulting in columns from the same table having closer representations. 
%closely represented columns of the unionable tables. 
However, column alignment requires contextual understanding specific to the column itself, which differs from other columns within a table. 
Worth noting, Starmie (bipartite), which matches columns of a table pair, is generally outperformed by Starmie (holistic). 
Our holistic matching approach considers a broader context by aligning a set of tables together, resulting in improved performance. 
The SANTOS benchmark contains a higher proportion of numerical columns compared to other benchmarks, which are not effectively embedded by Starmie. As a result, the holistic approach tends to create distinct clusters for numerical columns, leading to a decrease in recall and subsequently, the F1 score.
In conclusion, our holistic approach consistently achieves superior column alignments than bipartite matching with well-embedded columns. Accordingly, \ourmethod uses the best-performing Column-level RoBERTa model. 
% \revision{
\subsubsection{Efficiency}
\label{section:column_alignment_efficiency}
Next, we report runtime to align the columns in each benchmark. Note that this
%Column Alignment 
time depends on the time taken by the model to embed the columns and run the clustering algorithm. Since the embedding time for each model is significantly fast (on average, less than a second per column in our benchmarks) and they are insignificantly different, Column Alignment time is dominated by the clustering time. On average, the column alignment time per query is reasonable (35, 46, and 24 seconds in TUS-Sampled, SANTOS, and UGEN-V1 Benchmarks respectively).
% }
\subsection{Tuple Representation Evaluation}
\label{section:tuple_representation_evaluation}

% \begin{table}
% \small
% \setlength{\tabcolsep}{3pt}
% \caption{Effectiveness of different models in representing unionable tuples. The best score is shown in bold and the second best score is \underline{underlined}.}
% \caption{Effectiveness of unionable tuple representation models. %The b
% Best score is %shown in 
% bolded and the second best %score 
% is \underline{underlined}.}
% \begin{tabular}{l|cccc|cc}
% \toprule
%     &BERT&RoBERTa&sBERT&Ditto&DUST&DUST\\
%     &&&&&(BERT)&(RoBERTa)\\\hline
%     \textbf{Accuracy}&0.50&0.50&0.56&0.66&\underline{0.84}&\textbf{0.85}\\ 
% \bottomrule
% \end{tabular}
% \label{tab:tuple_representation_results}
% \end{table}

% \begin{table}
% \small
% \caption{Effectiveness of different models in representing unionable tuples. The best score is shown in bold and the second best score is \underline{underlined}.}
% \begin{tabular}{l|r}
% \toprule
%     \textbf{Model}&\makecell{\textbf{Accuracy}}\\ 
% \hline
%     BERT&0.50\\
%     RoBERTa&0.50\\
%     sBERT&0.56\\
%     Ditto&0.66\\
% \hline
%     DUST (BERT)&\underline{0.84}\\
%     DUST (RoBERTa)&\textbf{0.85}\\
% \bottomrule
% \end{tabular}
% \label{tab:tuple_representation_results}
% \end{table}

% \begin{figure}[t]
%     \centering
% \includegraphics[scale=.5]{figures/tuple_representation_results.pdf}
%     \caption{Effectiveness of different models in representing unionable tuples} 
%     \label{fig:tuple_representation_results}
% \end{figure}

Now, we compare the performance of DUST tuple representation against state-of-the-art baselines. 

\begin{figure}[h]
  \includegraphics[scale = 0.42]{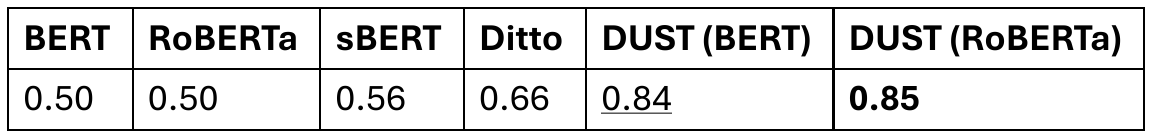}
  \caption{Unionable tuple representation Accuracy. %The b
Best score is %shown in 
bolded and the second best %score 
is \underline{underlined}.}
\label{tab:tuple_representation_results}
\end{figure}

\subsubsection{Experimental Setup and Evaluation Metric} 
\label{section:ex_setup_and_eval_metric}
To evaluate 
% the performance of 
\ourmethod's embedding model against other embedding models, we report \textbf{Accuracy} computed over the test set of TUS Fine-tuning Benchmark~(\cref{section:main_benchmarks}).
Each data point is a pair of tuples and an embedding model returns embeddings for each of them. We then compute their cosine distance. 
% \revision{
If the cosine distance is less than a threshold, we consider the tuple pairs predicted to be unionable. Otherwise, the tuple pairs are predicted to be non-unionable.
% } 
Based on empirical evaluation, we use a threshold of 0.7 which gives the best accuracy in the validation set.
We compute Accuracy using Scikit-learn's implementation.\footnote{\url{https://scikit-learn.org/stable/modules/model_evaluation.html}}
Mathematically, let $TP$, $FP$, $TN$, $FN$ be the number of true positives (unionable tuple pairs predicted as unionable), false positives (non-unionable tuple pairs predicated as unionable), true negative (non-unionable tuple pairs predicted as non-unionable) and false negative (non-unionable tuple pairs predicated as non-unionable). Then, 
% the $accuracy =\frac{TP + TN}{TP + TN + FP + FN}$.%is:
% calculated as:
%aamod: space saving
\begin{equation}
    Accuracy = \frac{TP + TN}{TP + TN + FP + FN}
\end{equation}

\subsubsection{Baselines} We now explain the baselines that we use for the experiments. Since there are no models for embedding unionable tuples, we adapt transformer-based embedding models that are successful in other semantic tasks.
Specifically, we use pre-trained \textbf{BERT}~\cite{2019_devlin_bert} and \textbf{RoBERTa}~\cite{2019_liu_roberta} models available in huggingface package.\footnote{\label{footnote:huggingface}\url{https://huggingface.co/models}}
Furthermore, Sentence BERT (\textbf{sBERT})~\cite{2019_reimers_sbert} is a recent state-of-the-art transformer-based model that is fine-tuned to closely embed sentences having similar meanings. We use sBERT as another baseline 
%as it is successful in understanding similar sentences and could 
to test if it can be fine-tuned to understand unionable tuples.$^{\ref{footnote:huggingface}}$
For consistency, we use the same serialization for all the models as we explained in~\cref{section:model_architecture}. 
Furthermore, entity matching is also a similar classification task where the objective is to test whether the two input tuples are about the same real-world entity. So, to evaluate if entity matching techniques can understand tuple unionability, we also compare the \ourmethod model against \textbf{Ditto}, a transformer-based model fine-tuned for the entity matching task~\cite{2020_li_ditto}. We implement Ditto using their public code.\footnote{\url{https://github.com/megagonlabs/ditto}}

\subsubsection{\ourmethod models}
We build two variations of \ourmethod. First we fine-tune BERT~\cite{2019_devlin_bert} (\textbf{DUST (BERT)}). Next, we fine-tune RoBERTa~\cite{2019_liu_roberta} (\textbf{DUST (RoBERTa)}). Note that both variations use the same serialization and the same fine-tuning architecture as described in~\cref{section:model_architecture}. We keep embedding dimension equal to 768 which is consistent with the pre-trained models. We train each model for at most 100 epochs. However, to avoid model overfitting, we apply an early-stopping strategy~\cite{2020_zhou_patience} with a patience of 10. Specifically, we compute the validation loss after each epoch and if the loss does not improve for 10 epochs, we terminate the training process and consider the model to be converged. 
% \revision{
It took around 30 hours to fine-tune both \textbf{DUST (BERT)} and \textbf{DUST (RoBERTa)}) using our experimental setup (four 8 GB GPUs).
Note that for dynamic data lakes, this fine-tuning can be optimized using more modern hardware and applied periodically when a lake has changed significantly.
%is an offline step that needs to be done only once. 
We share our fine-tuned models and training parameters.$^{\ref{footnote:dust_github}}$
% }

\subsubsection{Effectiveness}
\label{section:dust_embedding_effectiveness}
% We now report the effectiveness of the two \ourmethod embedding models against the baselines.
We report the accuracy of each baseline and the two variations of \ourmethod in~\cref{tab:tuple_representation_results}.
%First, we compare the performance of \ourmethod model's two variations. It is seen that RoBERTa-based 
\ourmethod (RoBERTa) performs slightly better than its BERT-based variation. A possible reason for this is that it is pre-trained over a larger number of parameters than BERT and has larger prior knowledge. Consequently, we use \ourmethod (RoBERTa) for all our experiments and to compare against other baselines. 
Next, we compare \ourmethod's performance against the baselines. It is seen that DUST outperforms the best baseline, Ditto, by over 15\% showing that embedding tuples to understand unionability is different than embedding them for entity matching. Moreover, \ourmethod outperforms sBERT by over 28\% indicating that  models that are good at detecting similar sentences can not capture intrinsic tabular properties and hence, they are not effective in determining tuple unionability. 
Further, we observe that both the pre-trained models, BERT and RoBERTa, perform as well as only tossing a coin, and are not able to understand tuple unionability. 
So, the low effectiveness of pre-trained models and fine-tuned models for embedding tuples for other tasks empirically validate the need to build \ourmethod model to embed unionable tuples.
% \revision{
Also, DUST embeddings are robust towards the change in column position within tuples, which we validate empirically in 
\paperorreport{the technical report~\cite{dust_technical_report}}{~\cref{section:appendix_column_variation_robustness}}.
% }

% \subsection{Tuple Diversification Evaluation}
\subsection{Tuple Diversification Experiments}
\label{section:tuple_diversification_evaluation}

\begin{table}
% \small
\fontsize{8}{10}\selectfont
\setlength{\tabcolsep}{1.5pt}
\centering
\caption{Reporting (i) the number of query tables for which each diversification algorithm gives the best Average and Min diversity scores and (ii) Average Time taken per Query by each algorithm in each benchmark.}
  \label{tab:tuple_diversity_results}
  \begin{tabular}{l|ccc|ccc}
    \toprule
    \multirow{2}{*}{\makecell[l]{\textbf{Method}}}& \multicolumn{3}{ c| }{\textbf{SANTOS}}& \multicolumn{3}{ c }{\textbf{UGEN-V1}}\\
    &\multicolumn{1}{c}{\textbf{\# Average}}
    &\multicolumn{1}{c}{\textbf{\# Min}}
    &\multicolumn{1}{c|}{\textbf{Time (s)}}
    &\multicolumn{1}{c}{\textbf{\# Average}}
    &\multicolumn{1}{c}{\textbf{\# Min}}
    &\multicolumn{1}{c}{\textbf{Time (s)}}\\
    \hline
    GMC&\textit{23}&\textit{1}&556&3&2&\textbf{<1}\\
    GNE&-&-&-&0&0&81\\
    CLT&0&0&\textbf{82}&\textit{18}&\textit{12}&\textbf{<1}\\
    \ourmethod&\textbf{27}&\textbf{49}&\textit{85}&\textbf{27}&\textbf{34}&\textbf{<1}\\
  \bottomrule
\end{tabular}
\end{table}

% \begin{figure*}[ht]
%   \includegraphics[scale = 0.42]{figures/santos_avg_diversity_cosine.pdf}
%   \caption{Average Diversity Score using Cosine distance (normalized from 0 to 1) in SANTOS Benchmark}
% \label{fig:santos_avg_diversity_cosine}
% \end{figure*}
%aamod: moved to TR.

% \begin{table}
% \small
% %\fontsize{8}{10}\selectfont
% \setlength{\tabcolsep}{2pt}
% \centering
% \caption{Average time (in seconds) per Query Table taken by each algorithm when diversifying tuples.}
%   \label{tab:tuple_diversity_time}
%   \begin{tabular}{l|ccc}
%     \toprule
%     \textbf{Method}&\textbf{TUS-Sampled}&\textbf{SANTOS}&\textbf{UGEN-V1}\\
%     \hline
%     GNE&-&-&81\\
%     GMC&483&556&<1\\
%     CLT&74&79&<1\\
%     \ourmethod&60&80&<1\\
%   \bottomrule
% \end{tabular}
% \end{table}

% \begin{table}
% \small
% %\fontsize{8}{10}\selectfont
% \setlength{\tabcolsep}{2pt}
% \centering
% \caption{Average time (in seconds) per Query Table taken by each algorithm when diversifying tuples.}
%   \label{tab:tuple_diversity_time}
%   \begin{tabular}{l|ccc}
%     \toprule
%     \textbf{Method}&\textbf{SANTOS}&\textbf{UGEN-V1}\\
%     \hline
%     GMC&556&<1\\
%     GNE&-&81\\
%     CLT&79&<1\\
%     \ourmethod&80&<1\\
%   \bottomrule
% \end{tabular}
% \end{table}

% }
Next, we evaluate the performance of our novel tuple diversification algorithm against state-of-the-art baselines (RQ2).
% \aamod{Update this section with new numbers.}
%aamod: done!
% Now we consider RQ2.
\subsubsection{Experimental Setup and Evaluation Metrics}
\label{section:tuple_diversification_evaluation_metrics}

As we highlighted in \cref{sec:div_eval}, we use two tuple diversification measures, namely, Average Diversity and Min Diversity. 
To keep the distance function consistent with the distance function that we use in the Tuple Representation Model (see~\cref{section:tuple_representation} and \cref{section:tuple_representation_evaluation}), we use cosine distance\footnote{\url{https://scikit-learn.org/stable/modules/generated/sklearn.metrics.pairwise.cosine_distances.html}} to measure the distance between the tuples throughout the experiments. Note that experiments with other distances such as Manhattan distance and Euclidean Distance are available in the \ourmethod github. With both distances, the relative performance of all the baselines is similar to using the cosine distance. 
% \revision{
In addition, we run an analysis to select $p$ in~\cref{alg:diversification}. We measure the improvement in the diversity metrics~(\cref{sec:div_eval}) when we increase the value of p, i.e., when we increase the number of candidate data lake tuples. Since, the improvement in the diversity metrics is either negative or insignificant for $p$ more than 2, we select $p= 2$ in our experiments. We provide further details in~\paperorreport{ the technical report~\cite{dust_technical_report}}{~\cref{section:appendix_p_threshold_analysis}}.
% }.

\subsubsection{Baselines} 
% \revision{
We compare the \ourmethod diversification algorithm with several state-of-the-art baselines from IR. 
% }
Specifically, \citet{2011_vieira_divdb} proposed different diversification algorithms and they have been adopted by a recent work~\cite{2023_mirzaei_tus_preference} to search for tables having diverse sets of columns. So, we  compare with the best-performing algorithms presented by \citet{2011_vieira_divdb}.
For each baseline, we use the default parameters suggested in the respective papers. We reproduced all algorithms following the original papers.

\introparagraph{GMC~\cite{2011_vieira_divdb}} 
GMC (Greedy Marginal Contribution) is used to diversify search results.
%the results in information retrieval literature.
It considers a trade-off between relevance and diversity based on a user's preference and formulates a diversity scoring function. Then, it greedily selects items one after another to add to the result set. An item is selected if it increases the diversity score computed using the items currently in the result set.

\introparagraph{GNE~\cite{2011_vieira_divdb}} GNE (Greedy Randomized with Neighborhood Expansion) is a randomized version of GMC algorithm. It starts by creating a candidate diverse set following the same diversity scoring function as GMC. Then it iterates over the candidate set to replace a random item with other items not included in the candidate set.

\introparagraph{CLT~\cite{2009_leuken_clt}} Since we use a clustering approach in our algorithm to retrieve candidate diverse data lake tuples, we also compare against a simple clustering-based baseline that was used to select diverse images. CLT generates k clusters and selects an item from each cluster using different strategies. To keep experiments consistent, we select each cluster's medoid in the diverse set. Further, we use the same clustering technique as ours with the same parameters.

\subsubsection{Effectiveness}
\label{section:tuple_div_effectiveness_report}
We run experiments with k = 100 in the SANTOS benchmark, and k = 30 in the UGEN-V1 benchmark. For each benchmark, we experiment with $s$ 
% \revision{
(number of candidate unionable tuples)
% }
equal to at most 2500. This selection is based on the number of unionable tuples per query available in the data lake of each benchmark and also to accommodate baselines that do not scale for large k and s. Nevertheless, for efficiency experiments, we will report results for larger k using a synthetic dataset~(\cref{section:tuple_diversification_efficiency}).
Note that the TUS benchmark is not used in this experiment as it was used to build the \ourmethod model.

Our benchmarks have multiple query tables and experiments on each query are independent of one another. So, following the literature~\cite{2011_vieira_divdb}, we report Average Diversity and Min Diversity Scores by each baseline on each query table separately.
Specifically, we report the number of queries for which each technique gives the highest Average Diversity and the highest Min Diversity Scores in comparison to other baselines~(\cref{tab:tuple_diversity_results}). The actual scores are provided in the github.$^{\ref{footnote:dust_github}}$

% \revision{

We first run %our 
an experiment using a simple \textbf{random} baseline, sampling %. We randomly sampled 
\textbf{k} tuples for each query and calculating their diversity scores. For a comprehensive analysis, we generated five random sets for each query using different random seeds and selected the highest-performing random set for each metric for comparison against DUST. In the SANTOS benchmark, DUST outperformed the random baseline in \emph{46 out of 50} queries for Average Diversity and in \emph{all} but one query for Min Diversity. In the UGEN benchmark (smaller tables),
random is better than DUST only for less than 25\% of queries for Average Diversity and in no queries for Min Diversity. These results show that random sampling is ineffective for tuple diversification. %Next, w
We now compare %DUST 
against the baselines from the literature.
% }

\ourmethod achieves the best Average Diversity scores for over 54 \% of the queries in the SANTOS benchmark, outperforming the second best method, GMC, by 8 \% (\cref{tab:tuple_diversity_results}). \ourmethod is the best method for more than half queries in UGEN-V1 benchmark as well, where the second-best method, CLT achieves the best performance for around 36\% of queries. 
% \revision{
Interestingly, for all queries in the SANTOS benchmark where GMC performs the best, and for over 88 \% of queries in the UGEN-V1 benchmark where CLT performs the best, \ourmethod is the second-best method which shows that \ourmethod is more effective than the baselines in diversifying the tuples. 
As GNE does not scale to large datasets, we evaluate it only in the UGEN-V1 benchmark where it is outperformed by all the baselines. In terms of Min Diversity, \ourmethod %continues to 
gives the best performance for almost all queries in the SANTOS benchmark and around 70\% of queries in the UGEN-V1 benchmark.
% } 
This shows that %our method of 
first selecting candidate data lake tuples and then ranking them based on their distance from the query tuples improves diversification.

\subsubsection{Efficiency}
\label{section:tuple_diversification_efficiency}

\begin{figure}
    {
    \centering
    \begin{minipage}[t]{0.5\textwidth}
    % \hspace{1.5cm}
    \centering\includegraphics[width=.5\linewidth]{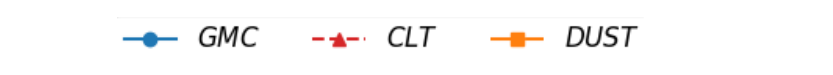}
    \end{minipage}%
    }
    % \centering
    \subfloat[Runtime vs \# of tuples]
    {
    \begin{minipage}[t]{0.45\linewidth}
    \includegraphics[width=\linewidth]{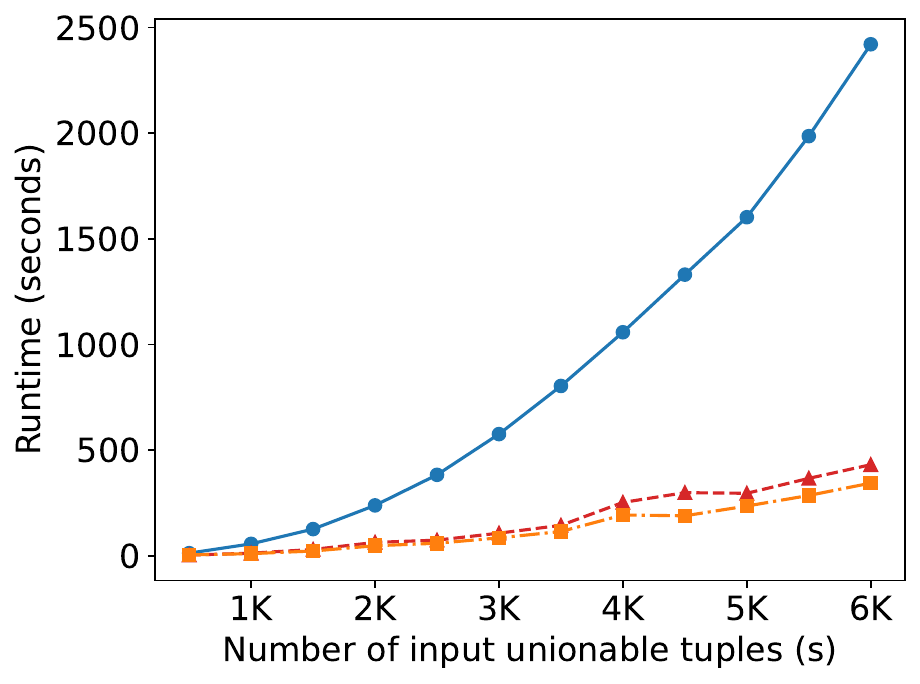}
    \end{minipage}%
    }
    % \centering
    \subfloat[Runtime vs k (s=5K)]{
    \begin{minipage}[t]{0.45\linewidth}
    \includegraphics[width=\linewidth]{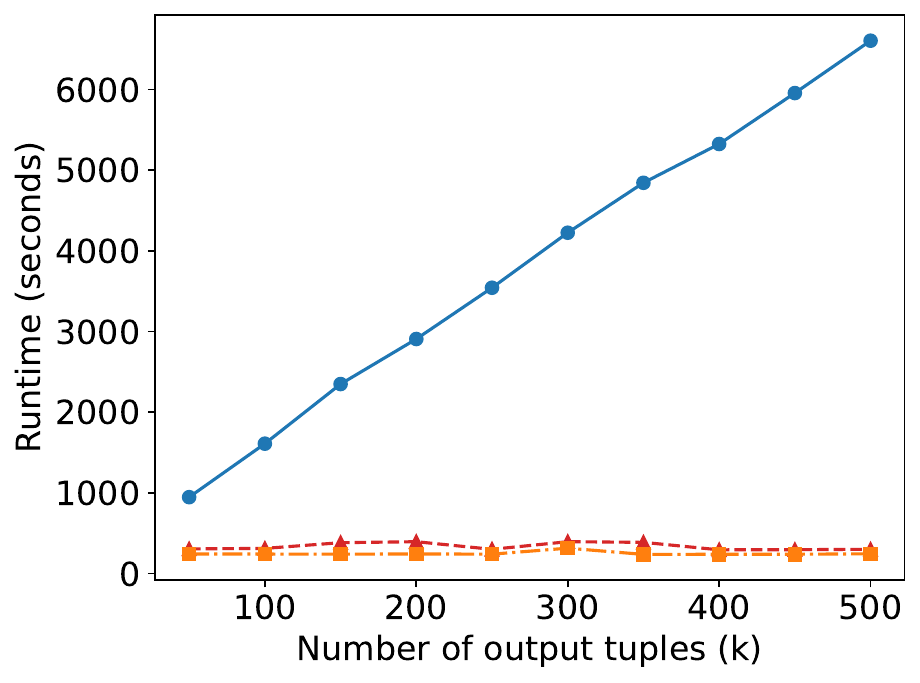}
    \end{minipage}%
    }
    \caption{\ourmethod Diversification Runtime against baselines}
    \label{fig:diversification_runtime}
\end{figure}
Now we compare \ourmethod's scalability against the baselines. In the SANTOS benchmark, \ourmethod is over 7 times faster than GMC and almost as fast as CLT on average
(\cref{tab:tuple_diversity_results}). 
In the UGEN-V1 benchmark, which has fewer unionable tuples per query than SANTOS, all but the GNE method diversify tuples within a second per query  on average. The GNE algorithm, which first generates a candidate diverse set and then improves it iteratively, is much slower than all the baselines. Hence, in 
%both 
%\rjm{do you mean all?  there are 3} 
%ak: changed to all
all the benchmarks, \ourmethod gives the best effectiveness being much faster than the second-best baseline (GMC) and as fast as other baseline (CLT).

To understand how the number of input unionable tuples (s) and the number of output tuples (k) impacts the runtime, we experiment with a query table and a variable number of tuples that are unionable with the query table. Specifically, we first vary s for constant k (k = 100) and report the runtime~(\cref{fig:diversification_runtime} (a)). We saw that the increase in the number of input tuples increases the runtime for GMC quadratically whereas \ourmethod is the fastest for which the runtime is linear to s with a very small slope. Also, when we vary k for s = 5000, and observe its impact on the runtime~(\cref{fig:diversification_runtime} (b)), it is seen that \ourmethod is not impacted by the increase in k. Furthermore, although \ourmethod has an additional step after clustering where it selects the unionable data lake tuples based on their distance from each query tuple, it has a similar runtime as a clustering baseline (CLT). Hence, \ourmethod's post-clustering phase is scalable practically. An analysis of pruning, showing its importance in reducing runtime, is provided 
in \paperorreport{the report~\cite{dust_technical_report}}{\cref{section:appendix_pruning_influence}}.
% DUST takes 990 seconds on average for each query, which drops to 85 seconds after pruning, without hurting DUST's effectiveness against the baselines

\subsection{DUST Against Table Search Techniques}
\label{section:tuple_search_evaluations}
%To understand if diversification of unionable tuples is useful, and how well \ourmethod diversifies searched unionable tuples, w
% \revision{
We now compare \ourmethod against two state-of-the-art table union search techniques, along with an LLM approach to this problem.
% }
%s~\cite{2023_fan_starmie} in terms of diversification. 
%Specifically, w
We report diversity scores (see~\cref{section:tuple_diversification_evaluation_metrics}) of k-tuples that \ourmethod outputs against that of k-tuples returned by table search technique.
% We measure Average Diversity and Min Diversity (see~\cref{section:tuple_diversification_evaluation_metrics}).
\subsubsection{Baselines} We now describe our baselines.

% \revision{
\introparagraph{D3L~\cite{2020_bogatu_d3l}}
$D3L$ aggregates different column unionability signals, such as value match, word embedding match, regular expressions, and so on, to search for the top-k unionable tables from a data lake. We implement D3L using its public code
% ~\cite{url_d3l}. 
.\footnote{\url{https://github.com/alex-bogatu/d3l}}
% }

\introparagraph{Starmie~\cite{2023_fan_starmie}}
Starmie is the state-of-the-art 
table union search technique that returns a set of top-k unionable tables, given a query table. To adopt Starmie for searching k unionable tuples, we index each tuple in the data lake as a separate table and search for the top-k tables. As each data lake contains a single tuple, the tuples from the top-k searched tables are the k output tuples.

\introparagraph{\textbf{LLM}~\cite{2020_floridi_gpt3}} 
We %also 
use a Large Language Model (GPT-3) as a baseline to generate diverse unionable tuples. %Specifically, w
We input query table tuples to GPT-3 and ask it to generate a set of k diverse unionable tuples with the given query table. The robustness and effectiveness of LLMs is impacted by the selected prompt and thus we have evaluated several prompts (the %used 
prompt is provided in \paperorreport{the github.$^{\ref{footnote:dust_github}}$}{~\cref{section:appendix_llm_baseline_prompt}}).

For a fair comparison with \ourmethod, we embed the output tuples by each baseline using \ourmethod embeddings and compute diversity scores over them.
We implement Starmie
% and Ver~\footnote{\url{https://github.com/TheDataStation/ver}}
using its public code with the default parameters.\footnote{\url{https://github.com/megagonlabs/starmie}}
We implement GPT-3 using its API~\cite{url_openai_chatgpt}.
% \footnote{\url{https://openai.com/chatgpt}}
% \revision{
Additionally, we implemented Ver~\cite{2023_gong_ver}, a Query-By-Example system that takes a small number of tuples as input and adds new data lake tuples to it.
\footnote{\url{https://github.com/TheDataStation/ver}} 
%identifies the subset of columns 
However, due to the creation of large indexes, the experiment for VER could not be scaled across any benchmarks. For instance, in UGEN-V1 Benchmark with 1000 data lake tables with a total of 8K columns and 10k rows~(\cref{fig:main_benchmarks}), Ver was automatically terminated after two days. 
% }

\subsubsection{Effectiveness}

\begin{table}
\small
%\fontsize{8}{10}\selectfont
\setlength{\tabcolsep}{2pt}
\centering
\caption{Number of query tables for which each diversification algorithm performs the best. 
}
  \label{tab:full_pipeline_effectiveness_results}
  \begin{tabular}{l|cc|cc}
    \toprule
    \multirow{2}{*}{\makecell[l]{\textbf{Method}}}& \multicolumn{2}{ c| }{\textbf{SANTOS}}& \multicolumn{2}{ c }{\textbf{UGEN-V1}}\\
    &\multicolumn{1}{c}{\textbf{\# Average}}
    &\multicolumn{1}{c}{\textbf{\# Min}}
    &\multicolumn{1}{c}{\textbf{\# Average}}
    &\multicolumn{1}{c}{\textbf{\# Min}}\\
    \hline
    Starmie&5&1&11&2\\
    LLM&-&-&\textit{14}&\textit{21}\\
    \ourmethod&\textbf{45}&\textbf{49}&\textbf{23}&\textbf{25}\\
  \bottomrule
\end{tabular}
\end{table}

% \begin{figure}
%   \includegraphics[scale = 0.45]{figures/full_pipeline_effectiveness_results.pdf}
%  \caption{Number of query tables for which each diversification algorithm performs the best.}
% \label{tab:full_pipeline_effectiveness_results}
% \end{figure}

We now report DUST's diversification effectiveness against baselines in SANTOS and UGEN-V1 benchmarks. We exclude LLM from the analysis in SANTOS benchmark as it was not scalable for the query tables with a large number of tuples.~\cref{tab:full_pipeline_effectiveness_results} reports the number of queries in each benchmark where each method performs the best. The actual diversity scores are provided in the github.$^{\ref{footnote:dust_github}}$ 
In the SANTOS benchmark, DUST achieves the highest Average Diversity score for around 90\% of queries and achieves the highest Min Diversity across almost all queries.
This can be attributed to the baseline's (Starmie's) inclination to favor similar tuples, resulting in the retrieval of tuples already present in the query table.
%, as elaborated in~\cref{example:union_search}. 
Further, in the UGEN-V1 benchmark, \ourmethod consistently yields the most diverse tuples for the highest number of queries, outperforming the LLM, the second best baseline, by around 18\% in terms of Average Diversity. Interestingly, for a given query, the LLM 
%is seen to 
generates a few diverse tuples but subsequently, it produces redundant ones. Nevertheless, \ourmethod could be scalable to search for 100s of tuples whereas LLM could not do so currently due to its token limits.
Also, Starmie, whose Mean Average Precision (MAP)~\cite{2023_khatiwada_santos} of searching for unionable tuples in UGEN-V1 is 64\%, shows better diversity performance than it does on the SANTOS Benchmark where its MAP is 78\%. Starmie finds more non-unionable tuples in UGEN-V1 (lower MAP), and such tuples are generally more diverse with each other.
These results support \ourmethod's necessity for diversifying unionable tuples to improve their usability. 
% for user analysis. 

% \revision{
% \subsubsection{Case Study}
\subsection{Case Study}
\label{section:case_study}

% \begin{figure*}
%   \includegraphics[scale = 0.55]{figures/imdb_case_study_full.pdf}
%   \caption{Number of novel values added to the different columns of query table by each method.}
% \label{fig:imdb_case_study}
% \end{figure*}
% \revision{
% \subsubsection{\textbf{IMDB Benchmark~\cite{url_imdb}}}

Finally, we show a case study, providing intuitive evidence of the benefits of diversification. %For a case study%~(\cref{section:case_study})
We have  created a small benchmark consisting of a query table and %its 
20 unionable tables derived from an IMDB movie dataset~\cite{url_imdb}. %We take an 
The original IMDB table %that 
contains information on nearly 500 recent movies, including title, director, genre, budget, filming location, language, %country of origin, 
and more. %Then, w
We sample its rows to create a query table and unionable tables. On average, the tables in this benchmark have 97 tuples and 13 columns (see repository)$^{\ref{footnote:dust_github}}$. Please note that this case study only aims to examine the diversity and thus only contains unionable tables/tuples.

\begin{figure}[h]
  \includegraphics[scale = 0.24]{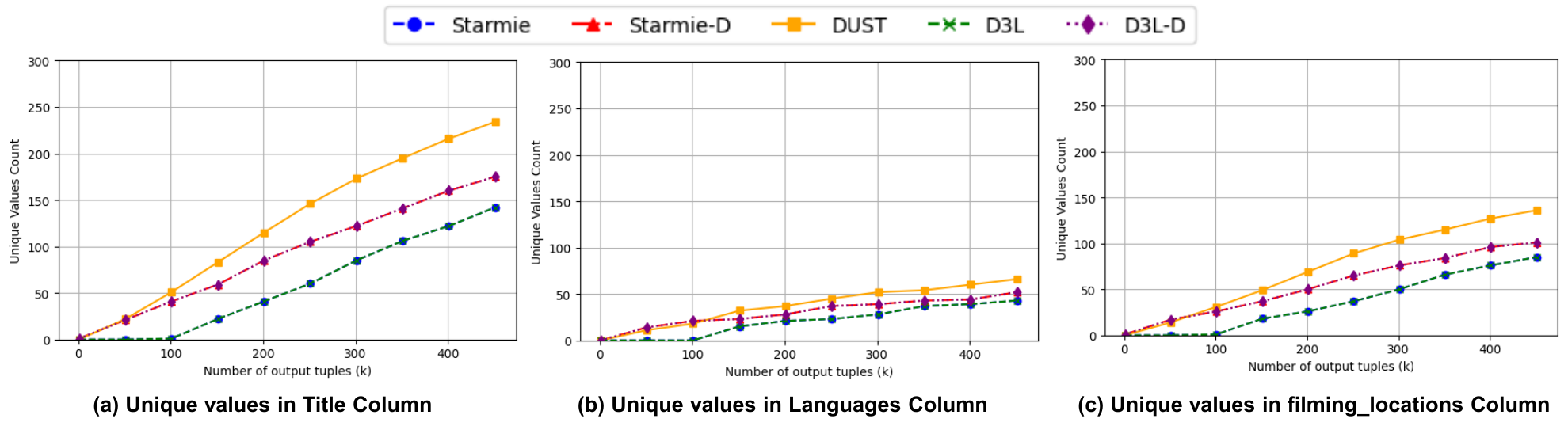}
  \caption{Number of novel values added to the different columns of query table by each method.}
\label{fig:imdb_case_study}
\end{figure}

% }

% \revision{
Using DUST, we search for a set of $k$ diverse tuples from this small  data lake.  We compare this to D3L and Starmie's result. Since both of the baselines output the top-N unionable tables, we (bag) union the output tables based on their ranking with the query table until we have a bag with at least $k$ tuples. From the resulting  unioned table, we select 
$k$ tuples (using SQL {\tt LIMIT k})  
%--excluding query tuples and duplicates--for comparison with 
and compare to the $k$ tuples produced by DUST. Starmie, D3L and other unionablity methods~\cite{2018_nargesian_tus,2023_khatiwada_santos,2023_hu_autotus} may find tables that overlap with the query table (and each other).
So we also examine a duplicates-free version of D3L and Starmie (dubbed D3L-D and Starmie-D, respectively, in \cref{fig:imdb_case_study}), in which we exclude duplicated tuples. Here, we take the set union of the top tables until this set contains at least $k$ tuples and again if there are more than $k$ we use the SQL {\tt LIMIT k} on this set. Then, we measure and report how many new values each method adds across different columns of the query table in IMDB Benchmark. %Please note that w
We refrain from comparing to an LLM %-based 
 baseline here as we aim to examine only movies from the given data lake.
 % }

% \revision{
In ~\cref{fig:imdb_case_study}, we plot the number of output tuples (k) on the X-axis and the number of unique values added by D3L, D3L-D, Starmie, Starmie-D, and DUST to the query table's (a) Title, (b) Languages, and (c) Filming\_locations columns on the Y-axis. Statistics for other columns are similar and available in the repository.$^{\ref{footnote:dust_github}}$
Our results show that DUST retrieves tuples with nearly 25\% more unique movie titles from the data lake than Starmie even when removing the duplicate tuples (Starmie-D). 
Interestingly, both D3L and Starmie add a similar number of unique values. This is because our evaluation focuses on a data lake composed exclusively of unionable tables. Since both baselines rely on table similarity to identify unionable candidates, their results tend to overlap as we discussed in~\cref{example:union_search}.   
% }
% }
% \revision{
To illustrate DUST's practical benefit even further, we also provide an anecdotal example showing the tuples it suggests against Starmie in \paperorreport{the report~\cite{dust_technical_report}}{\cref{section:appendix_anecdotal_example}}.
% }
\section{Conclusion}
\label{section:conclusion}
We introduced the problem of finding diverse unionable tuples from a data lake and presented \ourmethod as a first solution. We compared  each component of \ourmethod with relevant baselines and demonstrated its superior performance. We illustrated that \ourmethod outperforms 
% \revision{
two state-of-the-art table union search techniques and a Large Language Model
% } 
in terms of effectiveness and efficiency in discovering diverse unionable tuples.  Our work suggests the need for human-in-the-loop discovery techniques that allow a user to interactively select diverse tuples of interest to them.
% \revision{
We continue to explore augmenting our tuple embeddings together with additional "unembedded" tabular features %(e.g. 
such as column values and relationship between column pairs
% }
.%, and so on).}
%to make our system more robust towards outliers.}
\begin{acks}
This work was supported in part by NSF
under award numbers IIS-1956096, IIS-2107248, and IIS-2325632.   We acknowledge the support of the Canada Excellence Research Chairs (CERC) program. Nous remercions le Chaires d’excellence en recherche du Canada (CERC) de son soutien.
\end{acks}
\section*{Artifacts}
 The source code, data, and/or other artifacts have been made available at: \url{https://github.com/northeastern-datalab/dust}.

% \clearpage
\bibliographystyle{ACM-Reference-Format}
\bibliography{main}
\paperorreport{}{
\clearpage
\appendix
\section{Appendix}
\label{section:appendix}
\subsection{Preliminaries}
\label{section:appendix_preliminaries}

\subsubsection{Column Alignment Phase}
\label{section:appendix_column_alignment_phase}
We show the block diagram of Column Alignment Phase in~\cref{fig:column_alignment_illustration}. Specifically, after we form the cluster of columns, we discard the columns that do not consist the columns from the query table. The aligned columns are then used for further steps.

\subsection{Experiments}

\subsubsection{DUST Embedding Robustness towards Column Variation}
\label{section:appendix_column_variation_robustness}
% \revision{
During inference, the order of columns within two unionable tuples may vary, and it is essential for our model to remain robust to such changes. To evaluate this, we conducted an experiment using 18k tuples of the test set. For each tuple, we generated a shuffled version by randomly permuting the column order, then encoded both the original and shuffled versions using the DUST (RoBERTa) model. We computed the cosine similarity between the embeddings of the original and shuffled tuples. As illustrated in~\cref{fig:test_set_cosine_similarity_boxplot}, the similarity remains consistently high, averaging 0.98 with a standard deviation of 0.04. This demonstrates that DUST effectively maintains semantic consistency despite variations in column order.
% }

\subsubsection{Impact of $p$ in DUST Algorithm}
\label{section:appendix_p_threshold_analysis}
% \revision{
In tuple diversification, our goal is to select data lake tuples that contribute new and diverse information to the query table. This requires addressing two key factors. First, the selected new unionable tuples from the data lake should be diverse among themselves to avoid introducing redundancy within them when they are unioned to the query table. Second, even if these tuples are diverse among themselves, they may still be similar to those already present in the query table. To handle both cases, in~\cref{alg:diversification}, we first select more than $k$ candidate data lake tuples that are diverse among themselves. From this set, we then select $k$ tuples that are most diverse with respect to the query table tuples. We control the number of candidate data lake tuples using parameter p in~\cref{alg:diversification}. If $p$ is small, there's a higher likelihood that a candidate data lake tuple is already present in the query table but is still chosen due to a lack of alternative options. 
Conversely, a larger value of $p$ leads to more candidate data lake tuples for comparison against the query. However, this reduces the distances within candidate data lake tuples, and hence, reduces max-min diversity.

So to select $p$, we run an analysis where we see the improvement in the diversity metrics~(\cref{sec:div_eval}) on increasing the value of p on UGEN and SANTOS Benchmarks.
In~\cref{fig:p_threshold_analysis}, we report the value of $p$ on x-axis and the percentage improvement on y-axis. After $p=2$, the improvement in the diversity metrics is negative in terms of max-min diversity score and insignificant in terms of average diversity score. Specifically, as larger $p$ introduces more data lake tuples, their minimum distances reduces, impacting max-min diversification score. As there is no benefit of increasing $p$ beyond 2, we set it to 2 in our experiments.
% }

\begin{figure}
  \includegraphics[scale = 0.25]{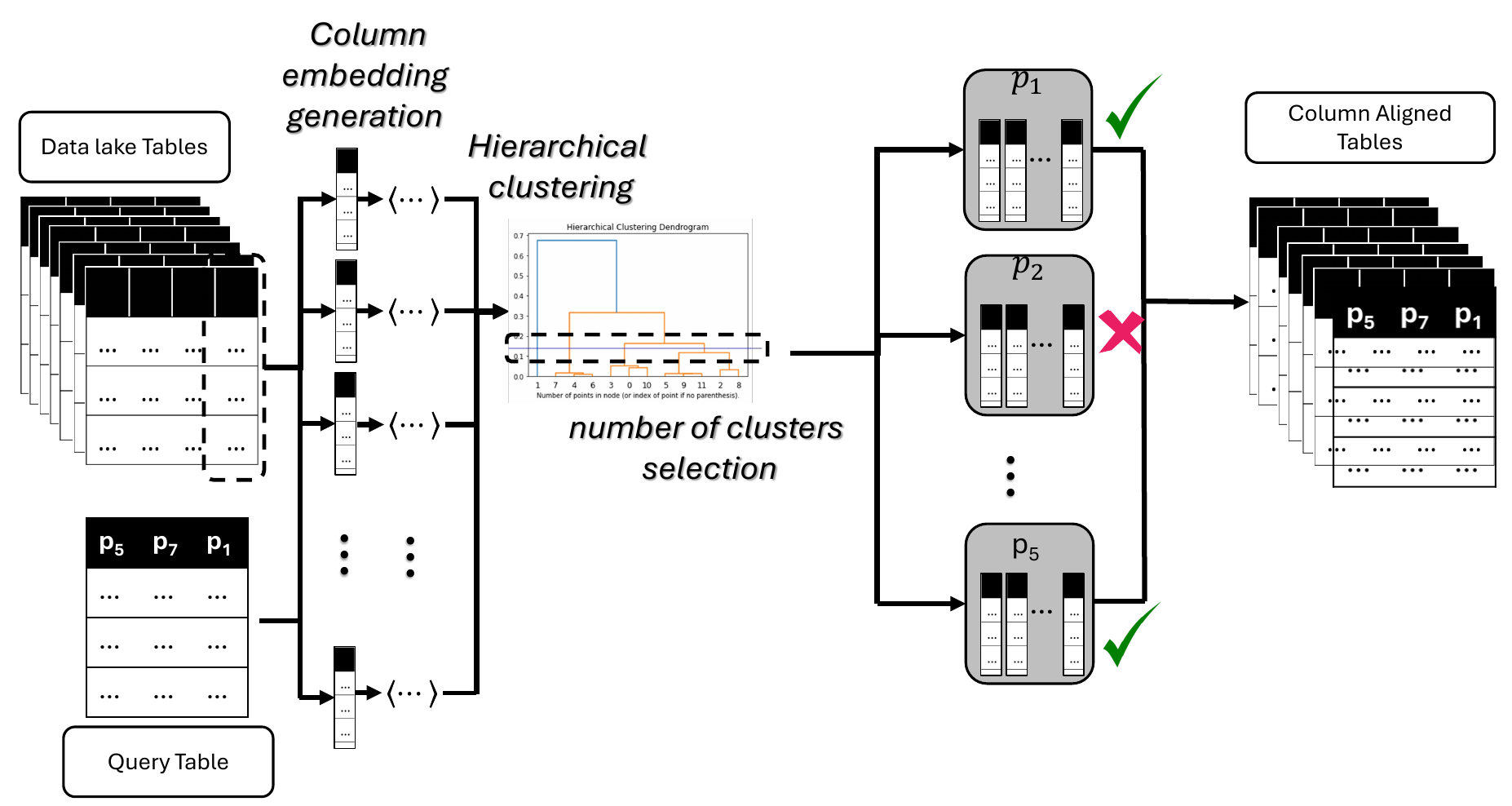}
  \caption{Aligning the Data lake table columns with the query table column. The columns that are not present in the query table are discarded.}
\label{fig:column_alignment_illustration}
\end{figure}

\begin{figure}
  \includegraphics[scale = 0.4]{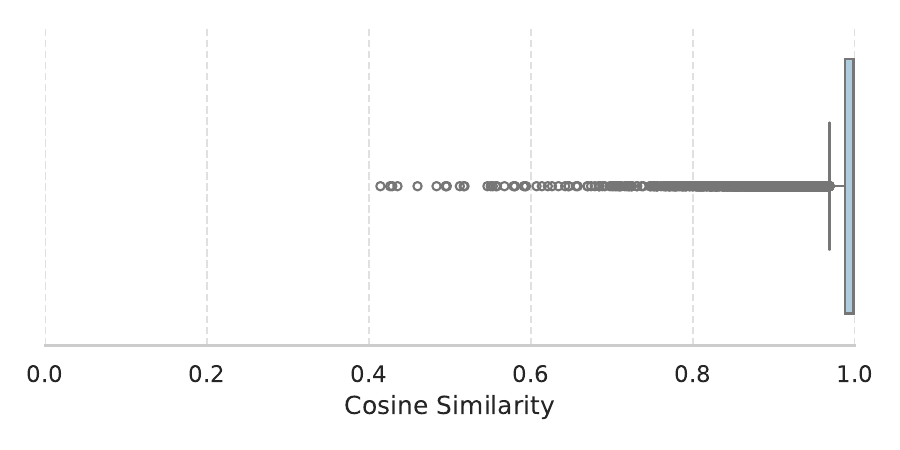}
  \caption{Distribution of cosine similarity between original and column-position shuffled tuples on Test set.}
\label{fig:test_set_cosine_similarity_boxplot}
\end{figure}

\begin{figure}
\centering
  \includegraphics[scale = 0.26]{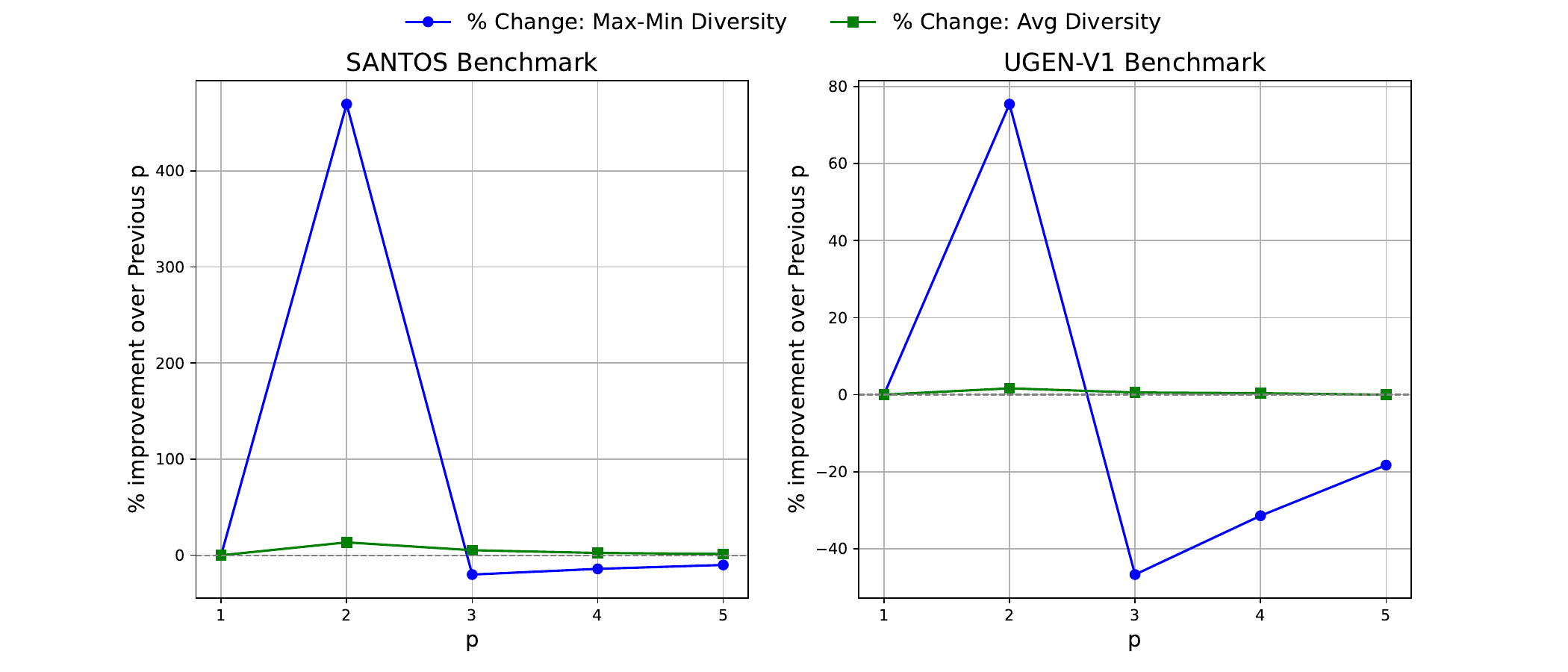}
  \caption{Impact of increasing $p$ in different benchmarks}
\label{fig:p_threshold_analysis}
\end{figure}

\begin{figure}
  \includegraphics[scale = 0.4]{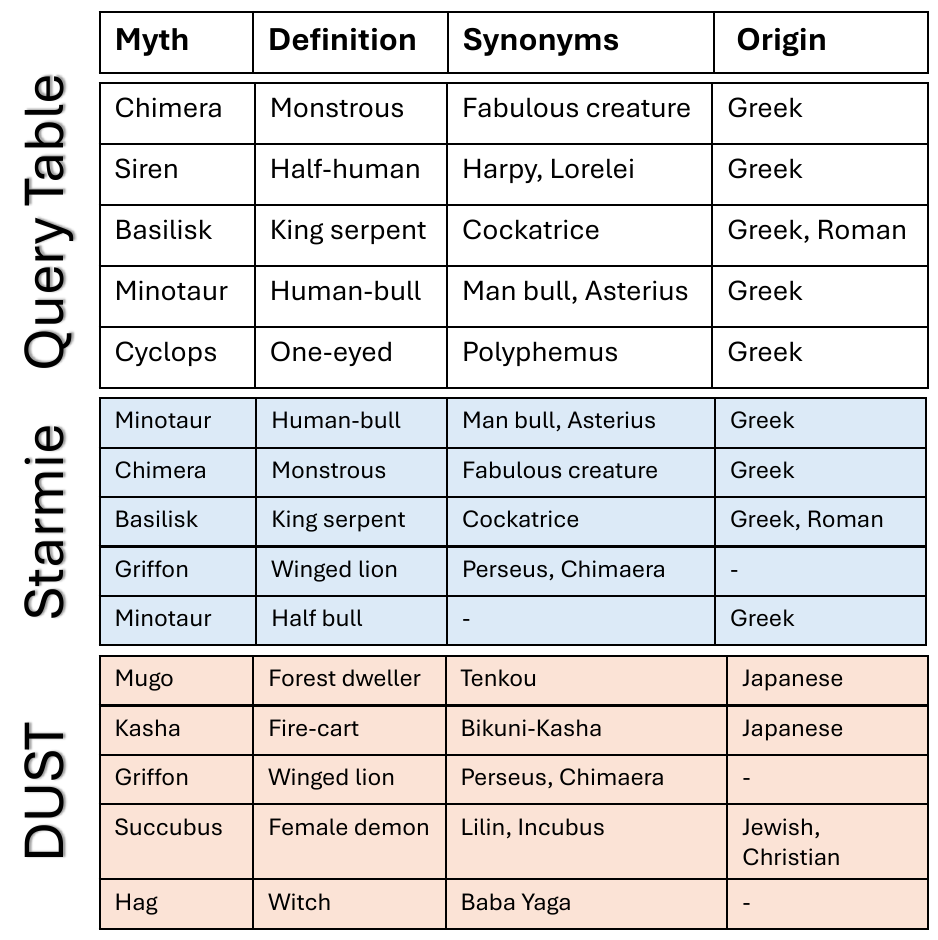}
  \caption{Tuples suggested by \ourmethod (bottom light red) and Starmie (middle light blue) for a given set of query table tuples (top)}
\label{fig:results_example}
\end{figure}

\subsubsection{Pre-diversification Pruning Influence}
\label{section:appendix_pruning_influence}
Recall that we prune candidate data lake tuples before we input them for diversification (see~\cref{sec:prun}). So, here we measure its impact during diversification empirically.
To allow a fair comparison, the pruning step (Section~\ref{sec:prun}) is applied for all baselines. To understand its importance, we analyze the runtime over the SANTOS Benchmark with and without pruning. We start with at most 10k unionable tuples and prune them to s = 2500 tuples, which we input for clustering. Without pruning, DUST takes 990 seconds on average for each query, which drops to 85 seconds after pruning, without hurting DUST's effectiveness against the baselines. The repo contains effectiveness results without pruning and for different values of s.$^{\ref{footnote:dust_github}}$

\subsubsection{LLM Baseline Prompt}
\label{section:appendix_llm_baseline_prompt}
The robustness and effectiveness of LLMs is impacted by the selected prompt. So, after trying multiple prompt configurations, we select the following prompt where we replace \{Table\} and \{k\} with the query table and the number of output tuples respectively.

\resultbox{
    \texttt{Given the following query table: \{Table\} \\
Generate \{k\} new tuples that are unionable to the query table. 
The generated tuples should be non-redundant and diverse with respect to the existing tuples. Return the tuples in pipe-separated format as the query table.}
}

\subsubsection{Diversification Anecdotal Example from Experiments}
\label{section:appendix_anecdotal_example}
% \revision{
To illustrate 
%the benefit of 
%aamod: space saving
\ourmethod{}'s benefit
in practice, we provide an anecdotal example showing the tuples it suggests against the baseline Starmie. The top part of \cref{fig:results_example} shows a subset of 5 tuples (out of 13) from the \texttt{Mythology} query table in UGEN-V1\footnote{\url{https://github.com/northeastern-datalab/gen/blob/main/data/ben_x/query/Mythology_5IDTY430.csv}}. We then show a subset of the top-5 (out of 30) tuples returned by Starmie and \ourmethod. As discussed in the introduction~(\cref{example:union_search}), Starmie suffers from redundancy. Starmie's top-3 tuples (\texttt{Minotaur}, \texttt{Chimera} and \texttt{Basilisk}) add no new information to the query table. Moreover, the fifth tuple, while different than the first, also refers to \texttt{Minotaur}. On the other end, \ourmethod{} returns new information with respect to the query table. Notably, the origins are also more diverse (not only \texttt{Greek/Roman}) showing the value of diversity also in the exact content on the tuples.
% }

}
\end{document}